%% file: BPH-16-003_temp.tex
\pdfoutput=1

\documentclass[11pt,twoside,a4paper,cmspaper,final,collab]{cms-tdr}
\def\svnVersion{481495P}\def\cmsCernNoTag{CERN-EP-2018-224}\def\cmsCernDate{\today}\def\cmsMessage{Published in the European Physical Journal C as \href{http://dx.doi.org/10.1140/epjc/s10052-018-6390-z}{\doi{10.1140/epjc/s10052-018-6390-z.}}}

\begin{document}\cmsNoteHeader{BPH-16-003}

\hyphenation{had-ron-i-za-tion}
\hyphenation{cal-or-i-me-ter}
\hyphenation{de-vices}
\RCS$Revision: 474542 $
\RCS$HeadURL: svn+ssh://svn.cern.ch/reps/tdr2/papers/BPH-16-003/trunk/BPH-16-003.tex $
\RCS$Id: BPH-16-003.tex 474542 2018-09-10 20:03:41Z alverson $

\ifthenelse{\boolean{cms@external}}{\providecommand{\cmsTable}[1]{\resizebox{\columnwidth}{!}{#1}}}{\providecommand{\cmsTable}[1]{#1}}
\ifthenelse{\boolean{cms@external}}{\providecommand{\tableCMS}[1]{#1}}{\providecommand{\tableCMS}[1]{\resizebox{\columnwidth}{!}{#1}}}
\providecommand{\NA}{\ensuremath{\text{---}}}
\newlength\cmsTabSkip\setlength{\cmsTabSkip}{1ex}
\newcommand{\mupm}      {\ensuremath{\mu^\pm}\xspace}
\newcommand{\mumu}      {\ensuremath{\mu^+\mu^-}\xspace}
\newcommand{\pip}       {\ensuremath{\pi^+}\xspace}
\newcommand{\pim}       {\ensuremath{\pi^-}\xspace}
\newcommand{\pipi}      {\ensuremath{\pi^+\pi^-}\xspace}
\newcommand{\pipm}      {\ensuremath{\pi^\pm}\xspace}
\newcommand{\kaon}      {\ensuremath{\PK}\xspace}
\newcommand{\Kp}        {\ensuremath{\kaon^+}\xspace}
\newcommand{\Km}        {\ensuremath{\kaon^-}\xspace}
\newcommand{\Kpm}       {\ensuremath{\kaon^\pm}\xspace}
\newcommand{\KS}        {\ensuremath{\kaon^0_{\mathrm{\scriptscriptstyle S}}}\xspace}
\newcommand{\Kstarz}    {\ensuremath{\kaon^{*0}}\xspace}
\newcommand{\Kstarzm}   {\ensuremath{\kaon^{*}(892)^0}\xspace}
\newcommand{\jpsi}      {\ensuremath{\cmsSymbolFace{J}\hspace{-.08em}/\hspace{-.14em}\psi}\xspace}
\newcommand{\B}         {\ensuremath{\cmsSymbolFace{B}}\xspace}
\newcommand{\Bu}        {\ensuremath{\B^+}\xspace}
\newcommand{\Bd}        {\ensuremath{\B^0}\xspace}
\newcommand{\Bs}        {\ensuremath{\B^0_\cPqs}\xspace}
\newcommand{\Bdst}      {\ensuremath{\B^{*0}}\xspace}
\newcommand{\Bust}      {\ensuremath{\B^{*+}}\xspace}
\newcommand{\Bst}       {\ensuremath{\B^{*}}\xspace}
\newcommand{\Buorst}    {\ensuremath{\B^{(*)+}}\xspace}
\newcommand{\Bdorst}    {\ensuremath{\B^{(*)0}}\xspace}
\newcommand{\Bone}      {\ensuremath{\B_{1}}\xspace}
\newcommand{\Bonem}     {\ensuremath{\B_{1}(5721)^0}\xspace}
\newcommand{\Btwost}    {\ensuremath{\B^{*}_{2}}\xspace}
\newcommand{\Btwostm}   {\ensuremath{\B^{*}_{2}(5747)^0}\xspace}
\newcommand{\Bsone}     {\ensuremath{\B_{\cPqs1}}\xspace}
\newcommand{\Bsonem}    {\ensuremath{\B_{\cPqs1}(5830)^0}\xspace}
\newcommand{\Bstwost}   {\ensuremath{\B^{*}_{\cPqs2}}\xspace}
\newcommand{\Bstwostm}  {\ensuremath{\B^{*}_{\cPqs2}(5840)^0}\xspace}
\newcommand{\Bsonetwost}{\ensuremath{\B^{(*)}_{\cPqs1,2}}\xspace}
\newcommand{\dmonepm}   {\ensuremath{\Delta M_{\Bsone}^{\pm}}\xspace}
\newcommand{\dmoneze}   {\ensuremath{\Delta M_{\Bsone}^{0}}\xspace}
\newcommand{\dmtwopm}   {\ensuremath{\Delta M_{\Bstwost}^{\pm}}\xspace}
\newcommand{\dmtwoze}   {\ensuremath{\Delta M_{\Bstwost}^{0}}\xspace}
\newcommand{\dmtwopmeq} {\ensuremath{M(\Bstwost)-M_{\Bu}^{\mathrm{PDG}}-M_{\Km}^{\mathrm{PDG}}}\xspace}
\newcommand{\dmtwozeeq} {\ensuremath{M(\Bstwost)-M_{\Bd}^{\mathrm{PDG}}-M_{\KS}^{\mathrm{PDG}}}\xspace}
\newcommand{\dmonepmeq} {\ensuremath{M(\Bsone)-M_{\Bust}^{\mathrm{PDG}}-M_{\Km}^{\mathrm{PDG}}}\xspace}
\newcommand{\dmonezeeq} {\ensuremath{M(\Bsone)-M_{\Bdst}^{\mathrm{PDG}}-M_{\KS}^{\mathrm{PDG}}}\xspace}
\newcommand{\dmtwopmzeeq}{\ensuremath{M(\Bstwost)-M_{\B}^{\mathrm{PDG}}-M_{\kaon}^{\mathrm{PDG}}}\xspace}
\newcommand{\dmonepmzeeq}{\ensuremath{M(\Bsone)-M_{\Bst}^{\mathrm{PDG}}-M_{\kaon}^{\mathrm{PDG}}}\xspace}
\newcommand{\pp}        {\ensuremath{\mathrm{pp}}\xspace}
\newcommand{\KSPiPi}    {\ensuremath{\KS\to\pipi}\xspace}
\newcommand{\BuJpsiK}   {\ensuremath{\Bu\to\jpsi\Kp}\xspace}
\newcommand{\BdJpsiKPi} {\ensuremath{\Bd\to\jpsi\Kp\pim}\xspace}
\newcommand{\BdJpsiKst} {\ensuremath{\Bd\to\jpsi\Kstarz}\xspace}
\newcommand{\BdJpsiKstm}{\ensuremath{\Bd\to\jpsi\Kstarzm}\xspace}
\newcommand{\KstarKPi}  {\ensuremath{\Kstarz\to\Kp\pim}\xspace}
\newcommand{\BssBuKm}   {\ensuremath{\Bstwost\to\Bu\Km}\xspace}
\newcommand{\BssmBuKm}  {\ensuremath{\Bstwostm\to\Bu\Km}\xspace}
\newcommand{\BssBustKm} {\ensuremath{\Bstwost\to\Bust\Km}\xspace}
\newcommand{\BsoBustKm} {\ensuremath{\Bsone\to\Bust\Km}\xspace}
\newcommand{\BuKm}      {\ensuremath{\Bu\Km}\xspace}
\newcommand{\BK}        {\ensuremath{\B\kaon}\xspace}
\newcommand{\BuPi}      {\ensuremath{\Bu\pim}\xspace}
\newcommand{\BssBdKS}   {\ensuremath{\Bstwost\to\Bd\KS}\xspace}
\newcommand{\BssmBdKS}  {\ensuremath{\Bstwostm\to\Bd\KS}\xspace}
\newcommand{\BssBdstKS} {\ensuremath{\Bstwost\to\Bdst\KS}\xspace}
\newcommand{\BsoBdstKS} {\ensuremath{\Bsone\to\Bdst\KS}\xspace}
\newcommand{\BsomBdstKS}{\ensuremath{\Bsonem\to\Bdst\KS}\xspace}
\newcommand{\BdKS}      {\ensuremath{\Bd\KS}\xspace}
\newcommand{\mBd}       {\ensuremath{M_{\Bd}}\xspace}
\newcommand{\mBu}       {\ensuremath{M_{\Bu}}\xspace}
\newcommand{\mBdst}     {\ensuremath{M_{\Bdst}}\xspace}
\newcommand{\mBust}     {\ensuremath{M_{\Bust}}\xspace}
\newcommand{\mpB}       {\ensuremath{M_{\B}^{\mathrm{PDG}}}\xspace}
\newcommand{\mpBd}      {\ensuremath{M_{\Bd}^{\mathrm{PDG}}}\xspace}
\newcommand{\mpBu}      {\ensuremath{M_{\Bu}^{\mathrm{PDG}}}\xspace}
\newcommand{\mpBst}     {\ensuremath{M_{\Bst}^{\mathrm{PDG}}}\xspace}
\newcommand{\mpBdst}    {\ensuremath{M_{\Bdst}^{\mathrm{PDG}}}\xspace}
\newcommand{\mpBust}    {\ensuremath{M_{\Bust}^{\mathrm{PDG}}}\xspace}
\newcommand{\mBuKm}     {\ensuremath{m_{\Bu\Km}}\xspace}
\newcommand{\mBdKS}     {\ensuremath{m_{\Bd\KS}}\xspace}
\newcommand{\mBK}       {\ensuremath{m_{\BK}}\xspace}
\newcommand{\mBuPi}     {\ensuremath{m_{\Bu\pim}}\xspace}
\newcommand{\Gammabss}  {\ensuremath{\Gamma_{\Bstwost}}\xspace}
\newcommand{\sigmBd}    {\ensuremath{\sigma_{\mBd}}\xspace}
\newcommand{\sigmBu}    {\ensuremath{\sigma_{\mBu}}\xspace}
\newcommand{\ptmu}      {\ensuremath{\pt(\mupm)}\xspace}
\newcommand{\ptmumu}    {\ensuremath{\pt(\mumu)}\xspace}
\newcommand{\ptbu}      {\ensuremath{\pt(\Bu)}\xspace}
\newcommand{\ptbd}      {\ensuremath{\pt(\Bd)}\xspace}
\newcommand{\rron}      {\ensuremath{R^{0\pm}_{2}}\xspace}
\newcommand{\rrtw}      {\ensuremath{R^{0\pm}_{1}}\xspace}
\newcommand{\rrth}      {\ensuremath{R^{\pm}_{2*}}\xspace}
\newcommand{\rrfo}      {\ensuremath{R^{0}_{2*}}\xspace}
\newcommand{\rrfi}      {\ensuremath{R^{\pm}_{\sigma}}\xspace}
\newcommand{\rrsi}      {\ensuremath{R^{0}_{\sigma}}\xspace}

\cmsNoteHeader{BPH-16-003}

\title{Studies of $\Bstwostm$ and $\Bsonem$ mesons including the observation of the $\BssmBdKS$ decay
in proton-proton collisions at $\sqrt{s}=8\TeV$}
\titlerunning{Studies of $\Bstwostm$ and $\Bsonem$ mesons}

\date{\today}

\abstract{
Measurements of $\Bstwostm$ and $\Bsonem$ mesons are performed using a
data sample of proton-proton collisions corresponding to an integrated
luminosity of $19.6\fbinv$, collected with the CMS detector at the LHC
at a centre-of-mass energy of $8\TeV$.
The analysis studies $P$-wave $\Bs$ meson decays into
$\Buorst\Km$ and $\Bdorst\KS$, where the $\Bu$ and $\Bd$ mesons are
identified using the decays $\BuJpsiK$ and $\BdJpsiKstm$.
The masses of the $P$-wave $\Bs$ meson states are measured and the
natural width of the $\Bstwostm$ state is determined.
The first measurement of the mass difference between the charged and
neutral $\Bst$ mesons is also presented.
The $\Bstwostm$ decay to $\BdKS$ is observed, together with a
measurement of its branching fraction relative to the $\BssmBuKm$ decay.
}

\hypersetup{%
pdfauthor={CMS Collaboration},%
pdftitle={Studies of Bs*2(5848)0 and Bs1(5830)0 mesons including the observation of the Bs*2(5840)0 to B0 K0S decay in proton-proton collisions at sqrt(s) = 8 TeV},%
pdfsubject={CMS},%
pdfkeywords={CMS, physics, b hadrons, heavy flavour spectroscopy, hadron spectroscopy, experimental results}}

\maketitle

\section{Introduction}

The $P$-wave $\Bs$ states are the bound states of $\cPqb$ and $\cPqs$ quarks
with an orbital angular momentum $L=1$. Since the b quark is considerably
heavier than the strange quark, heavy-quark effective theory
(HQET)~\cite{Close:1992uv,Grozin:2004yc}
can be applied to describe this system. In the HQET framework, the state
can be described by $L$ and the spin of the
light quark, providing a total angular momentum
of the light subsystem $j=L\pm\frac{1}{2}$.
In the case of $L=1$, this results in $j=\frac{1}{2}$ or $j=\frac{3}{2}$.
Including the additional splitting from the spin of the heavy b quark
results in a total angular momentum $J=j\pm\frac{1}{2}$,
yielding two doublets, with the four states denoted as:
$\B^{*}_{\cPqs0}$ ($j=\frac{1}{2}$, $J^P=0^+$),
$\B^{*}_{\cPqs1}$ ($j=\frac{1}{2}$, $J^P=1^+$),
$\Bsone$ ($j=\frac{3}{2}$, $J^P=1^+$), and
$\Bstwost$ ($j=\frac{3}{2}$, $J^P=2^+$).
The two former states have not been observed to date, while the latter two
are known as the $\Bsonem$ and $\Bstwostm$ mesons, respectively.
For simplicity in this paper, shortened symbols are used
to denote the following particles:
$\Kstarz \equiv \Kstarzm$, $\Bone \equiv \Bonem$,
$\Btwost \equiv \Btwostm$, $\Bsone \equiv \Bsonem$,
$\Bstwost \equiv \Bstwostm$, and $\Bsonetwost$ refers to either $\Bsone$ or $\Bstwost$.
Charge-conjugate states are implied throughout the paper.
According to HQET, the decays $\BssBuKm$, $\BssBustKm$, and $\BsoBustKm$
are allowed and should proceed through a $D$-wave transition,
while the decay $\Bsone\to\Bu\Km$ is forbidden. Similar conclusions
apply to the decays into $\Bdorst\KS$.

Orbitally excited states of the $\Bs$ meson were observed by the CDF and D0
Collaborations via the decays into $\Buorst\Km$~\cite{Aaltonen:2007ah, Abazov:2007af}.
More recently, the LHCb Collaboration presented a more precise study of these
states and observed the decay $\Bstwostm\to\Bust\Km$~\cite{Aaij:2012uva},
favouring the spin-parity assignment $J^P=2^+$ for the $\Bstwostm$ state.
The CDF Collaboration subsequently presented a study of excited $\PB$ meson
states~\cite{Aaltonen:2013atp} that included measurements of the
$\Bsonetwost\to\Buorst\Km$ decays.
Table~\ref{tab:PreviousResults} summarizes all the available experimental
$\Bsonetwost$ results.

\begin{table*}[tbh]
  \renewcommand*{\arraystretch}{1.07}
\centering
\topcaption{Results on the masses, mass differences, and natural widths of
the $\Bsonetwost$ mesons from previous measurements. The mass differences
are defined as $\dmonepm\equiv\dmonepmeq$ and $\dmtwopm\equiv\dmtwopmeq$, where the
PDG superscript refers to the world-average mass values at the
time of each publication. \label{tab:PreviousResults}}
    \begin{tabular}{l|cccc}
                    & CDF~\cite{Aaltonen:2007ah}            & D0~\cite{Abazov:2007af}
                    & LHCb~\cite{Aaij:2012uva}              & CDF~\cite{Aaltonen:2013atp}      \\
\hline
$M(\Bstwost)\,[\MeV]$&$5839.6\pm0.7$    & $5839.6\pm1.3$    & $5839.99\pm0.21$  & $5839.7\pm0.2$    \\
$M(\Bsone)\,[\MeV]$ & $5829.4\pm0.7$    & $-$               & $5828.40\pm0.41$  & $5828.3\pm0.5$    \\
$\dmonepm\,[\MeV]$  & $10.73\pm0.25$    & $11.5\pm1.4$      & $10.46\pm0.06$    & $10.35\pm0.19$    \\
$\dmtwopm\,[\MeV]$  & $66.96\pm0.41$    & $66.7\pm1.1$      & $67.06\pm0.12$    & $66.73\pm0.19$    \\
[\cmsTabSkip]
$\Gamma(\Bstwost)\,[\MeV]$& \NA         & \NA               & $1.56\pm0.49$     & $1.4\pm0.4$       \\
$\Gamma(\Bsone)\,[\MeV]$&   \NA         & \NA               & \NA               & $0.5\pm0.4$
    \end{tabular}
  \renewcommand*{\arraystretch}{1.0}
\end{table*}

In this paper, the first observation of the $\BssBdKS$ decay and a measurement
of its branching fraction relative to that of the $\BssBuKm$ decay are presented.
The $\Bu$ and $\Bd$ candidates are reconstructed using the
$\Bu\to\jpsi(\mumu)\Kp$ and $\Bd\to\jpsi(\mumu)\Kstarz(\Kp\pim)$
decays, respectively.
Measurements of several ratios of branching fractions and ratios of production cross
sections times branching fractions are determined using the formulae:
\begin{linenomath}
\ifthenelse{\boolean{cms@external}}
{ 
\begin{multline}
\rron = \frac{\mathcal{B}(\BssBdKS)}{\mathcal{B}(\BssBuKm)} \\
=
\frac{N(\BssBdKS)}{N(\BssBuKm)}\,
\frac{\epsilon(\BssBuKm)}{\epsilon(\BssBdKS)}
\\
\times \frac{\mathcal{B}(\BuJpsiK)}{\mathcal{B}(\BdJpsiKst)\mathcal{B}(\KstarKPi)\mathcal{B}(\KSPiPi)}, \label{eqON}
\end{multline}
\begin{multline}
\rrtw = \frac{\mathcal{B}(\BsoBdstKS)}{\mathcal{B}(\BsoBustKm)} \\
=
\frac{N(\BsoBdstKS)}{N(\BsoBustKm)}\,\frac{\epsilon(\BsoBustKm)}{\epsilon(\BsoBdstKS)}\\
\times
\frac{\mathcal{B}(\BuJpsiK)}{\mathcal{B}(\BdJpsiKst)\mathcal{B}(\KstarKPi)\mathcal{B}(\KSPiPi)}, \label{eqTW}
\end{multline}
\begin{multline}
\rrth = \frac{\mathcal{B}(\BssBustKm)}{\mathcal{B}(\BssBuKm)}\\ =
\frac{N(\BssBustKm)}{N(\BssBuKm)} \,
\frac{\epsilon(\BssBuKm)}{\epsilon(\BssBustKm)}, \label{eqTH}
\end{multline}
\begin{multline}
\rrfo = \frac{\mathcal{B}(\BssBdstKS)}{\mathcal{B}(\BssBdKS)}\\ =
\frac{N(\BssBdstKS)}{N(\BssBdKS)} \,
\frac{\epsilon(\BssBdKS)}{\epsilon(\BssBdstKS)}, \label{eqFO}
\end{multline}
\begin{multline}
\rrfi = \frac{\sigma(\pp\to\Bsone \mathrm{X})\,\mathcal{B}(\BsoBustKm)}{\sigma(\pp\to\Bstwost \mathrm{X})\,\mathcal{B}(\BssBuKm)} \\
=
\frac{N(\BsoBustKm)}{N(\BssBuKm)} \,
\frac{\epsilon(\BssBuKm)}{\epsilon(\BsoBustKm)}, \label{eqFI}
\end{multline}
\begin{multline}
\rrsi = \frac{\sigma(\pp\to\Bsone\mathrm{X})\,\mathcal{B}(\BsoBdstKS)}{\sigma(\pp\to\Bstwost \mathrm{X})\,\mathcal{B}(\BssBdKS)} \\
=
\frac{N(\BsoBdstKS)}{N(\BssBdKS)} \,
\frac{\epsilon(\BssBdKS)}{\epsilon(\BsoBdstKS)}, \label{eqSI}
\end{multline}
} 
{ 
\begin{align}
\rron =& \frac{\mathcal{B}(\BssBdKS)}{\mathcal{B}(\BssBuKm)} =
\frac{N(\BssBdKS)}{N(\BssBuKm)} \,
\frac{\epsilon(\BssBuKm)}{\epsilon(\BssBdKS)} \nonumber \\
&\times \frac{\mathcal{B}(\BuJpsiK)}{\mathcal{B}(\BdJpsiKst)\mathcal{B}(\KstarKPi)\mathcal{B}(\KSPiPi)}, \label{eqON}\\
\rrtw =& \frac{\mathcal{B}(\BsoBdstKS)}{\mathcal{B}(\BsoBustKm)} =
\frac{N(\BsoBdstKS)}{N(\BsoBustKm)} \,
\frac{\epsilon(\BsoBustKm)}{\epsilon(\BsoBdstKS)}\nonumber\\
&\times\frac{\mathcal{B}(\BuJpsiK)}{\mathcal{B}(\BdJpsiKst)\mathcal{B}(\KstarKPi)\mathcal{B}(\KSPiPi)}, \label{eqTW} \\
\rrth =& \frac{\mathcal{B}(\BssBustKm)}{\mathcal{B}(\BssBuKm)} =
\frac{N(\BssBustKm)}{N(\BssBuKm)} \,
\frac{\epsilon(\BssBuKm)}{\epsilon(\BssBustKm)}, \label{eqTH}\\
\rrfo =& \frac{\mathcal{B}(\BssBdstKS)}{\mathcal{B}(\BssBdKS)} =
\frac{N(\BssBdstKS)}{N(\BssBdKS)} \,
\frac{\epsilon(\BssBdKS)}{\epsilon(\BssBdstKS)}, \label{eqFO} \\
\rrfi =& \frac{\sigma(\pp\to\Bsone \mathrm{X})\,\mathcal{B}(\BsoBustKm)}{\sigma(\pp\to\Bstwost \mathrm{X})\,\mathcal{B}(\BssBuKm)} =
\frac{N(\BsoBustKm)}{N(\BssBuKm)} \,
\frac{\epsilon(\BssBuKm)}{\epsilon(\BsoBustKm)}, \label{eqFI} \\
\rrsi =& \frac{\sigma(\pp\to\Bsone\mathrm{X})\,\mathcal{B}(\BsoBdstKS)}{\sigma(\pp\to\Bstwost \mathrm{X})\,\mathcal{B}(\BssBdKS)} =
\frac{N(\BsoBdstKS)}{N(\BssBdKS)} \,
\frac{\epsilon(\BssBdKS)}{\epsilon(\BsoBdstKS)}, \label{eqSI}
\end{align}} 
\end{linenomath}
where $\mathrm{X}$ stands for an inclusive reaction, and
$N(\mathrm{A}\to \mathrm{BC})$ and $\epsilon(\mathrm{A}\to \mathrm{BC})$
correspond to the number of
$\mathrm{A}\to \mathrm{BC}$ decays observed in data and the total efficiency
for the $\mathrm{A}\to \mathrm{BC}$ decay, respectively.
The branching fractions of the decays
$\Bust\to\Bu\gamma$ and $\Bdst\to\Bd\gamma$ are assumed to be 100\%.
Additionally, the mass differences in the studied decays and
the natural width of the $\Bstwostm$ state are measured, as well as the
mass differences $\mBd-\mBu$ and $\mBdst-\mBust$.
The data sample corresponds to an integrated luminosity of $19.6\fbinv$ of
proton-proton collisions at $\sqrt{s}= 8\TeV$, collected by
the CMS experiment~\cite{Chatrchyan:2008zzk} at the CERN LHC in 2012.

\section{The CMS detector}
The central feature of the CMS apparatus is a superconducting solenoid
of 6\unit{m} internal diameter, providing a magnetic field of 3.8\unit{T}.
Within the solenoid volume are a silicon pixel and strip tracker, a lead
tungstate crystal electromagnetic calorimeter, and a brass and
scintillator hadron calorimeter, each composed of a barrel and two
endcap sections. Muons are detected
in the pseudorapidity range  $\abs{\eta}<2.4$
in gas-ionization chambers embedded in
the steel flux-return yoke outside the solenoid.
The main subdetectors used for the present analysis are
the silicon tracker and the muon detection system.
The silicon tracker measures charged particles within the
range $\abs{\eta} < 2.5$.
For nonisolated particles with transverse momentum $1 < \pt < 10\GeV$
and $\abs{\eta} < 1.4$,
the track resolutions are typically 1.5\% in $\pt$ and 25--90\,(45--150)\mum
in the transverse (longitudinal) impact parameter~\cite{Chatrchyan:2014fea}.
Matching muons to tracks measured in the silicon tracker
results in a relative \pt resolution for muons
with $\pt < 10\GeV$ of 0.8--3.0\% depending on
$\abs{\eta}$~\cite{Chatrchyan:2012xi}.
A more detailed description of the CMS detector, together with a definition
of the coordinate system used and the relevant kinematic variables,
can be found in Ref.~\cite{Chatrchyan:2008zzk}.

Events of interest are selected using a two-tiered trigger
system~\cite{Khachatryan:2016bia}. The first level, composed of custom
hardware processors, uses information from the calorimeters and muon detectors
to select events at a rate of around 100\unit{kHz} within a time interval
of less than 4\mus.
The second level, known as the high-level trigger
(HLT), consists of a farm of processors running a version of the full event
reconstruction software optimized for fast processing, and reduces the event
rate to around 1\unit{kHz} before data storage.

\section{Event reconstruction and selection\label{sec:evesel}}

{\tolerance=800
The data sample is collected with an HLT algorithm
designed to select events with two muons consistent with originating
from a charmonium resonance decaying at a significant distance from the beam axis.
The requirements imposed at the trigger level include
$\ptmu>3.5\GeV$, $\abs{\eta(\mupm)}<2.2$, $\ptmumu>6.9\GeV$,
dimuon vertex $\chi^2$ fit probability $P_{\text{vtx}}(\mumu) > 10\%$,
dimuon invariant mass $1.0<M(\mumu)<4.8\GeV$,
distance between the beam axis and the reconstructed dimuon vertex position
in the transverse plane $L_{xy}(\mumu) > 3 \sigma_{L_{xy}(\mumu)}$,
where $\sigma_{L_{xy}(\mumu)}$ is the uncertainty in $L_{xy}(\mumu)$,
and the cosine of the dimuon candidate pointing angle to the beam axis
$\cos(\vec{L}_{xy}(\mumu)$, $\vec{\pt}(\mumu)) > 0.9$.
The pointing angle is the angle between the $\mumu$ candidate momentum in
the transverse ($x$--$y$) plane and the vector from the
beam axis position to the reconstructed dimuon vertex in the transverse plane.
\par}

{\tolerance=800
The reconstruction and selection of the $\PB$ meson candidates are similar
to those described in Ref.~\cite{Sirunyan:2017ofq}.
The analysis requires two muons of opposite charge
that must match those that triggered the event readout.
The trigger requirements are confirmed and the $\jpsi$ candidates are selected
by tightening the dimuon mass region to $[3.04,\,3.15]\GeV$.
\par}

\begin{figure}[tbh]
\centering
    \includegraphics[width=0.48\textwidth]{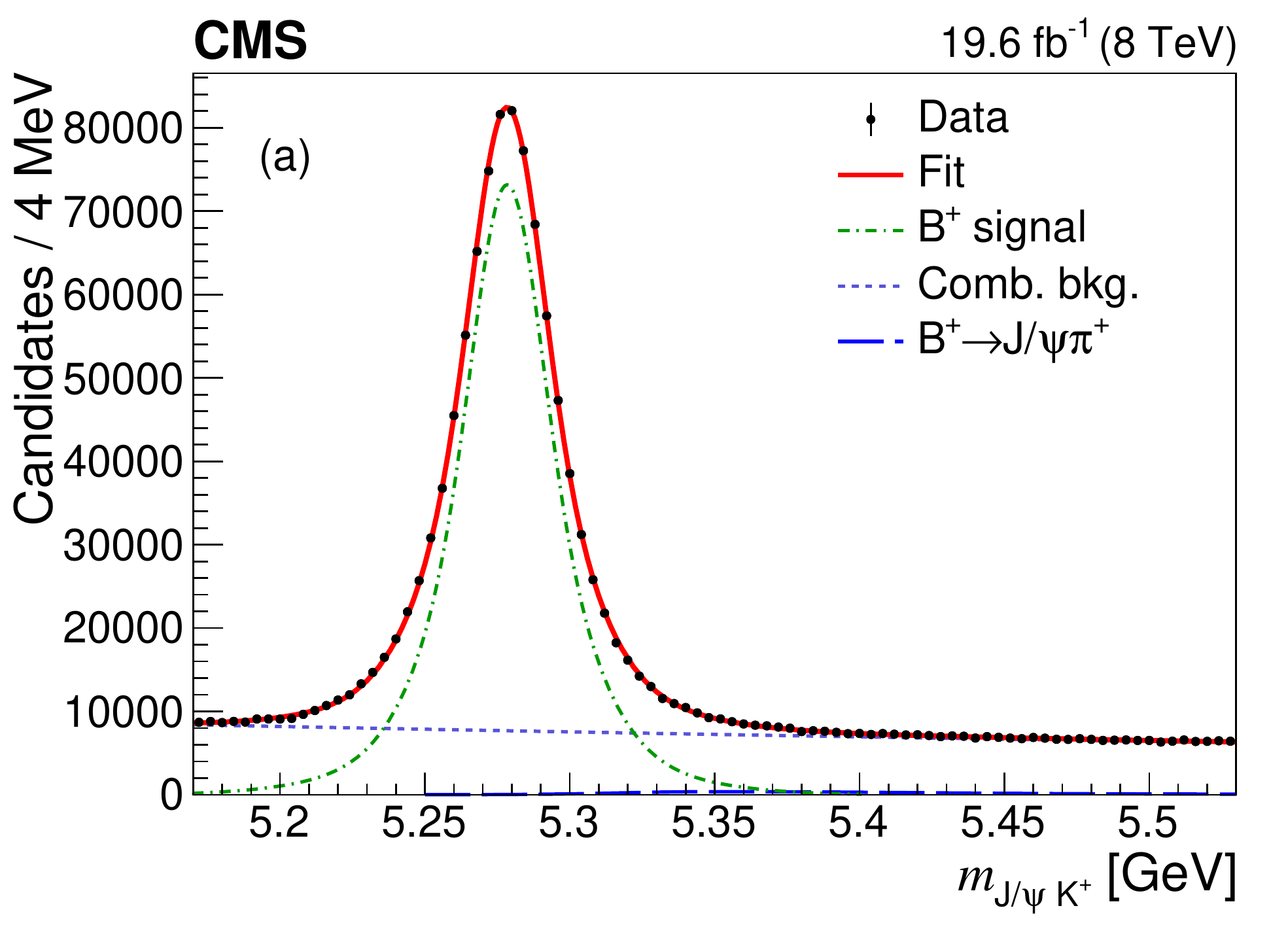}
    \includegraphics[width=0.48\textwidth]{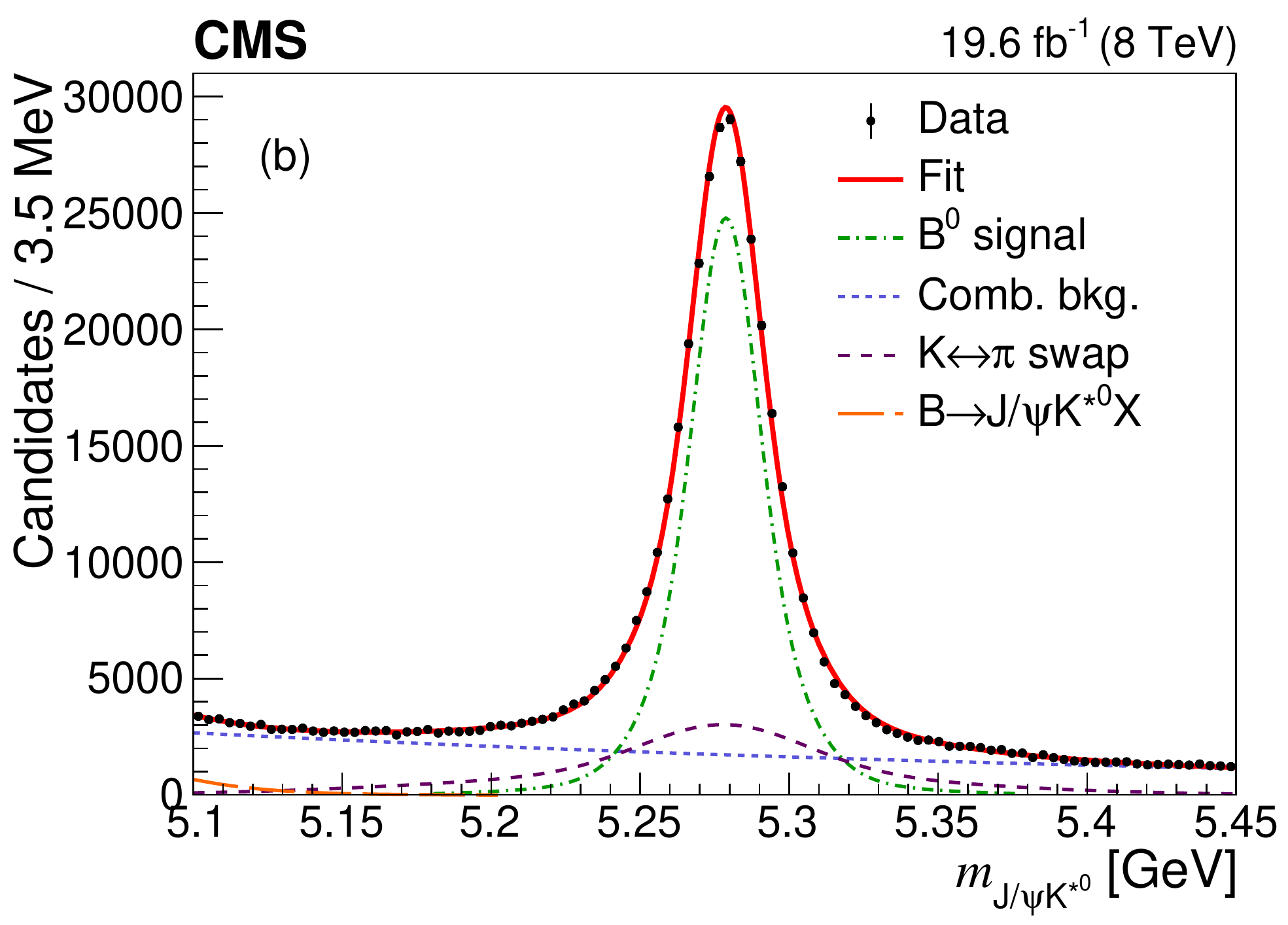}
\caption{
Invariant mass distributions of (a) $\jpsi\Kp$ and (b) $\jpsi\Kstarz$
candidates in data with the fit results superimposed.
The points represent the data, with the vertical bars giving the corresponding
statistical uncertainties. The thick curves are results of the fits,
the dash-dotted lines display the signal contributions, and the short-dashed lines show
the combinatorial background contributions.
The long-dashed line shows in (a) the contribution from the $\Bu\to\jpsi\pip$ decay, and
in (b) the contribution from partially reconstructed $\B\to\jpsi\Kstarz\mathrm{X}$ decays.
The dashed line in (b) displays the contribution from swapping
$\Kpm\to\pipm$ in the reconstruction. \label{fig:mBFit}
}
\end{figure}

{\tolerance=1200
The $\BuJpsiK$ candidates are constructed by combining the selected $\jpsi$
candidates with a track having $\pt>1\GeV$ to which the kaon mass is assigned.
The muon candidates must also satisfy the soft-muon identification
criteria described in Ref.~\cite{Chatrchyan:2012xi}, and
the kaon candidates must pass the high-purity track requirements
detailed in Ref.~\cite{Chatrchyan:2014fea}.
A kinematic fit to the three tracks is performed that constrains
the dimuon invariant mass to the world-average $\jpsi$ mass~\cite{PDG}.
From all the reconstructed \pp collision vertices in an event, the primary vertex (PV)
is chosen as the one with the smallest $\Bu$ pointing angle.
This pointing angle is the angle between the $\Bu$ candidate momentum and
the vector from the PV to the reconstructed $\Bu$ candidate vertex.
Furthermore, in this procedure, if any of the three tracks used in the
$\Bu$ candidate reconstruction are included in the fit of the chosen
PV, they are removed, and the PV is refitted.
The $\Bu$ candidates are required to have $\ptbu>10\GeV$,
$P_{\text{vtx}}(\Bu)>1\%$, $L_{xy}(\Bu) > 5 \sigma_{L_{xy}(\Bu)}$,
and $\cos(\vec{L}_{xy}(\Bu)$, $\vec{\pt}(\Bu))> 0.99$.
The invariant mass distribution of the $\BuJpsiK$ candidates
is shown in Fig.~\ref{fig:mBFit}(a).
An unbinned extended maximum-likelihood fit is performed to this
distribution using a triple-Gaussian function
with common mean for the signal,
an exponential function for the combinatorial background, and a fixed-shape function,
derived from simulation, accounting for the Cabibbo-suppressed $\Bu\to\jpsi\pip$ decay.
The parameters of the signal and the combinatorial background contributions, as well
as the yields of the different components, are free in the fit.
The effective resolution of the signal function ($\sigmBu$) found from
simulation of about 24\MeV is consistent with the resolution measured in data.
The invariant mass $M(\Bu)$ returned by the vertex fit is
required to lie in the range $[5.23, 5.33]\GeV$, corresponding to
a $\pm2\sigmBu$ window around the $\Bu$ mass.
\par}

The selected $\Bu$ candidates are combined with each track originating
from the chosen PV with the charged kaon mass assigned to it.
The track charge must be opposite to that of the reconstructed
$\Bu$ meson candidate (in the following, this track is referred to as $\Km$).
The kaon candidate is required to fulfill the standard high-purity track
requirements~\cite{Chatrchyan:2014fea} and have $\pt(\Km)>1\GeV$.

The reconstruction of $\Bd\to\jpsi(\mumu)\Kstarz(\Kp\pim)$ candidates
is similar to the one used for the charged decay mode.
The dimuon combinations forming $\jpsi$ candidates are
obtained using the same algorithm.
The $\Bd$ candidates are constructed from the selected $\jpsi$
candidates and two tracks of opposite charge, assumed to be from a
kaon and a pion. The tracks are required to satisfy standard high-purity
track requirements~\cite{Chatrchyan:2014fea} and have $\pt>1\GeV$. Those kaon and pion
candidates that can be matched to a signal in the muon chambers are rejected.

The $\Bd$ candidates are obtained by performing a kinematic vertex fit
to the four tracks described above that constrains the dimuon invariant mass
to that of the $\jpsi$ meson~\cite{PDG}.
The candidates are required to have $L_{xy}(\Bd) > 5 \sigma_{L_{xy}(\Bd)}$,
$P_{\text{vtx}}(\mumu\Kp\pim)>1\%$, $\cos(\vec{L}_{xy}(\Bd),\vec{\pt}(\Bd))> 0.99$,
and $\ptbd>10\GeV$.
To reject the contribution from $\Bs\to\jpsi\phi$ decay, the invariant
mass of the two hadron tracks, if both are assigned the kaon mass, is required
to be above 1.035\GeV.
We demand that the $\Kp\pim$ invariant mass is within 90\MeV of the $\Kstarz$
mass~\cite{PDG}.  If both the $\Kp\pim$ and $\Km\pip$ hypotheses pass this
selection, then the $\Kp\pim$ invariant mass must be closer to the $\Kstarz$
mass than the $\Km\pip$ invariant mass.
The invariant mass distribution of the selected $\BdJpsiKPi$ candidates
is shown in Fig.~\ref{fig:mBFit}(b).
It is fitted with a sum of a triple-Gaussian function with a common mean for
the signal, a double-Gaussian function accounting for the
$\Kpm\to\pipm$ swapped (KPS) component, where the second Gaussian is
asymmetric, and an exponential function for the combinatorial background.
An additional Gaussian function is included to account for the
partially reconstructed $\B\to\jpsi\Kstarz\mathrm{X}$ background near the left
edge of the fit region.
The resolution parameters of the signal function and the parameters of the
KPS are fixed to the values obtained in simulation; the other parameters
are free in the fit. The effective resolution of the signal
function ($\sigmBd$) found from the simulation is about 19\MeV.
The $\Bd$ candidate returned by the vertex fit is required to have
an invariant mass in the range 5.245 to 5.313\GeV, corresponding to approximately
$\pm2\sigmBd$ around the known $\Bd$ mass~\cite{PDG}.
The fit results are used to extract the fraction of the KPS with respect
to the signal yield in the $\Bd$ signal region of $(18.9\pm0.3)\%$, where
the uncertainty is statistical only.

{\tolerance=800
The selected $\Bd$ candidates are combined with \KS candidates that are formed
from detached two-prong vertices, assuming the decay $\KSPiPi$,
as described in Ref.~\cite{Khachatryan:2010pw}.
The two-pion invariant mass is required to be
within $\pm20\MeV$ of the \KS mass~\cite{PDG}, which corresponds
approximately to 4 times the \pipi mass resolution.
The two pion tracks are refitted with their invariant mass constrained
to the known \KS mass, and the obtained \KS candidate is required to satisfy
$P_{\text{vtx}}(\KS)>1\%$ and $\cos(\vec{L}_{xy}(\KS),\vec{\pt}(\KS))> 0.999$.
Multiple candidates from the same event are not removed.
\par}

Simulated events that are used to obtain relative efficiencies and
invariant mass resolutions are produced with
\PYTHIA v6.424~\cite{Sjostrand:2006za}.
The $\cPqb$ hadron decays are modelled with \EVTGEN 1.3.0~\cite{Lange2001152}.
Final-state photon radiation is included in \EVTGEN using
\PHOTOS~\cite{BARBERIO1991115,BARBERIO1994291}.
The events are then passed through a detailed \GEANTfour-based
simulation~\cite{Agostinelli2003250} of the CMS detector with the
same trigger and reconstruction algorithms as used for the data.
The simulation includes effects from multiple \pp interactions in the same
or nearby beam crossings (pileup) with the same multiplicity
distribution as observed in data.
Matching of the reconstructed candidates to the generated particles
is obtained by requiring $\Delta R=\sqrt{\smash[b]{(\Delta\eta)^2 + (\Delta\phi)^2}}$
to be $<$0.015 for \pipm and \Kpm, $<$0.004 for muons, and $<$0.020 for \KS,
where $\Delta\eta$ and $\Delta\phi$ are the differences in pseudorapidity
and azimuthal angle (in radians), respectively, between the
three-momenta of the reconstructed and generated particles.

\section{Fits to the \texorpdfstring{$\BK$}{BK} invariant mass distributions}
For every invariant mass distribution fit discussed in this section,
the functional models for the signal and the combinatorial background components
are chosen such that a good description of the binned distribution
is obtained. The description quality is verified using the difference between the
data and fit result, divided by the statistical uncertainty in the data and
also with $\chi^2$ tests.

\subsection{\texorpdfstring{$\BuKm$}{B+K-} invariant mass
 \label{sec:mBuKdata}}

To improve the $\BuKm$ invariant mass resolution, the variable $\mBuKm$ is computed as
\begin{equation*}
\mBuKm = M(\Bu\Km) - M(\Bu) + M_{\Bu}^{\mathrm{PDG}},
\end{equation*}
where $M(\Bu\Km)$ is the invariant mass of the reconstructed
\BuKm combination, $M(\Bu)$ is the reconstructed $\Bu$ mass, and
$M_{\Bu}^{\mathrm{PDG}}$ is the world-average $\Bu$ meson mass~\cite{PDG}.

\begin{figure}[tb]
  \centering
       \includegraphics[width=0.48\textwidth]{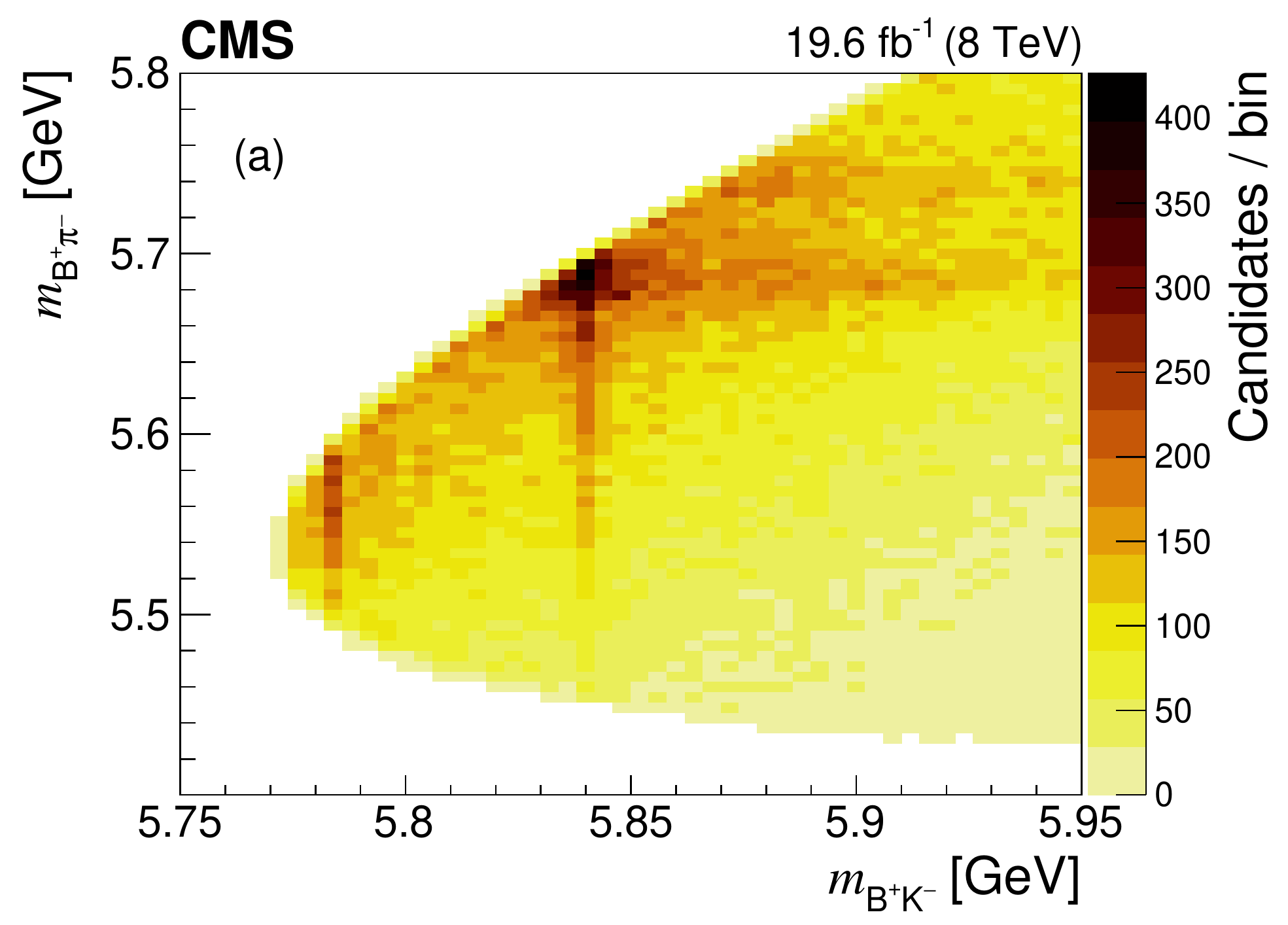}
       \includegraphics[width=0.48\textwidth]{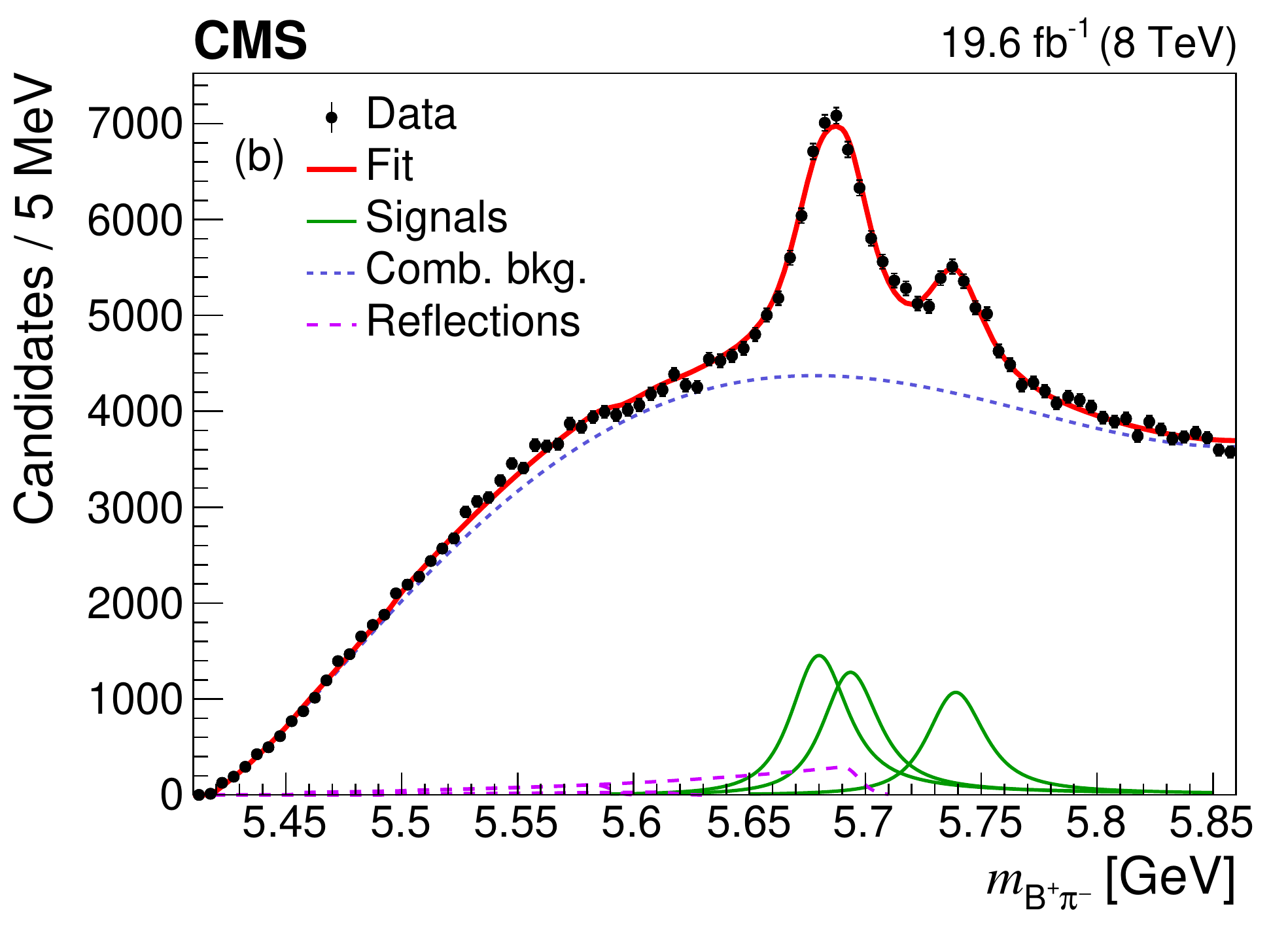}
  \caption{
(a) Two-dimensional distribution of $\mBuKm$ versus $\mBuPi$ in data.
(b) The fitted $\BuPi$ invariant mass distribution. The points represent the data,
the thick solid curve is the fit projection, the thin lines indicate the three
excited $\Bd$ signal contributions, the short-dashed curve is the combinatorial background,
and the long-dashed lines show the contributions from the excited $\Bs$ decays.
\label{fig:mBuPiFfits} }
\end{figure}

The decays of excited $\Bd$ mesons $\Bone\to\Bust\pim$,
$\Btwost\to\Bu\pim$, and $\Btwost\to\Bust\pim$ contribute to the
obtained $\BuKm$ mass distribution, as seen from the two-dimensional
distribution in Fig.~\ref{fig:mBuPiFfits}(a).
It is important to take into account these background contributions
in the fits to the $\mBuKm$ distribution.
Simulated samples of these decays are reconstructed
in the same way as the collision events to obtain the corresponding reflection shapes
in the $\mBuKm$ distribution. In order to measure the yields of these reflections,
the $\BuPi$ invariant mass, $\mBuPi$, is computed the same way as $\mBuKm$.
Fits are performed on the $\mBuPi$ distribution observed in data,
using the same data set, with a pion mass assigned to the track instead of a kaon mass.
Then the obtained yields of these contributions are used
in the fits to the $\mBuKm$ distribution.

{\tolerance=1000
The measured $\mBuPi$ distribution is presented in Fig.~\ref{fig:mBuPiFfits}(b).
Clear enhancements are seen around 5.65--5.75\GeV, corresponding to the
decays of excited $\Bd$ mesons.
An unbinned extended maximum-likelihood fit is performed to this distribution.
The three signal functions accounting for the $\Btwost\to\Bu\pim$,
$\Btwost\to\Bust\pim$, and $\Bone\to\Bust\pim$ decays are $D$-wave
relativistic Breit--Wigner (RBW) functions, convolved with a double-Gaussian
resolution function, with parameters fixed according to the simulation
(the typical effective resolution is about 5.5\MeV,
significantly below the natural widths of the states).
As verified in simulations, the signal shapes of $\Btwost\to\Bust\pim$
and $\Bone\to\Bust\pim$ decays (where the photon from the $\Bust$ decay is lost
and only the $\BuPi$ mass is reconstructed) are simply shifted by the mass
difference $\mpBust-\mpBu=45.34\pm0.23\MeV$~\cite{PDG}.
The combinatorial background is parametrized by the
function $(x-x_0)^{\alpha}\, P_n(x)$,
where $x\equiv\mBuPi$, $x_0$ is the threshold value, $\alpha$ is a free parameter,
and $P_n$ is a polynomial of degree $n$, with $n=3$.
Additional, relatively small contributions come from the
excited $\Bs$ decays.
They are included in the fit with free normalizations and fixed shapes,
obtained from the simulation.
\par}

In the nominal fit, the masses and natural widths of the excited $\Bd$ mesons are fixed
to their world-average values~\cite{PDG}.
The fit region is not extended to values above 5865\MeV
to avoid having to model the $\B(5970)$ contribution~\cite{Aaltonen:2013atp}.
The fitted event yields are about 8500, 10\,500, and 12\,000 for the
$\Btwost\to\Bu\pim$, $\Btwost\to\Bust\pim$, and $\Bone\to\Bust\pim$ signals,
respectively.

{\tolerance=800
Figure~\ref{fig:mBuKmFits}(a) shows the measured $\mBuKm$ distribution.
The three peaks from lower to higher mass correspond to the decays
$\BsoBustKm$, $\BssBustKm$, and $\BssBuKm$.
\begin{figure}[tb]
  \centering
    \includegraphics[width=0.48\textwidth]{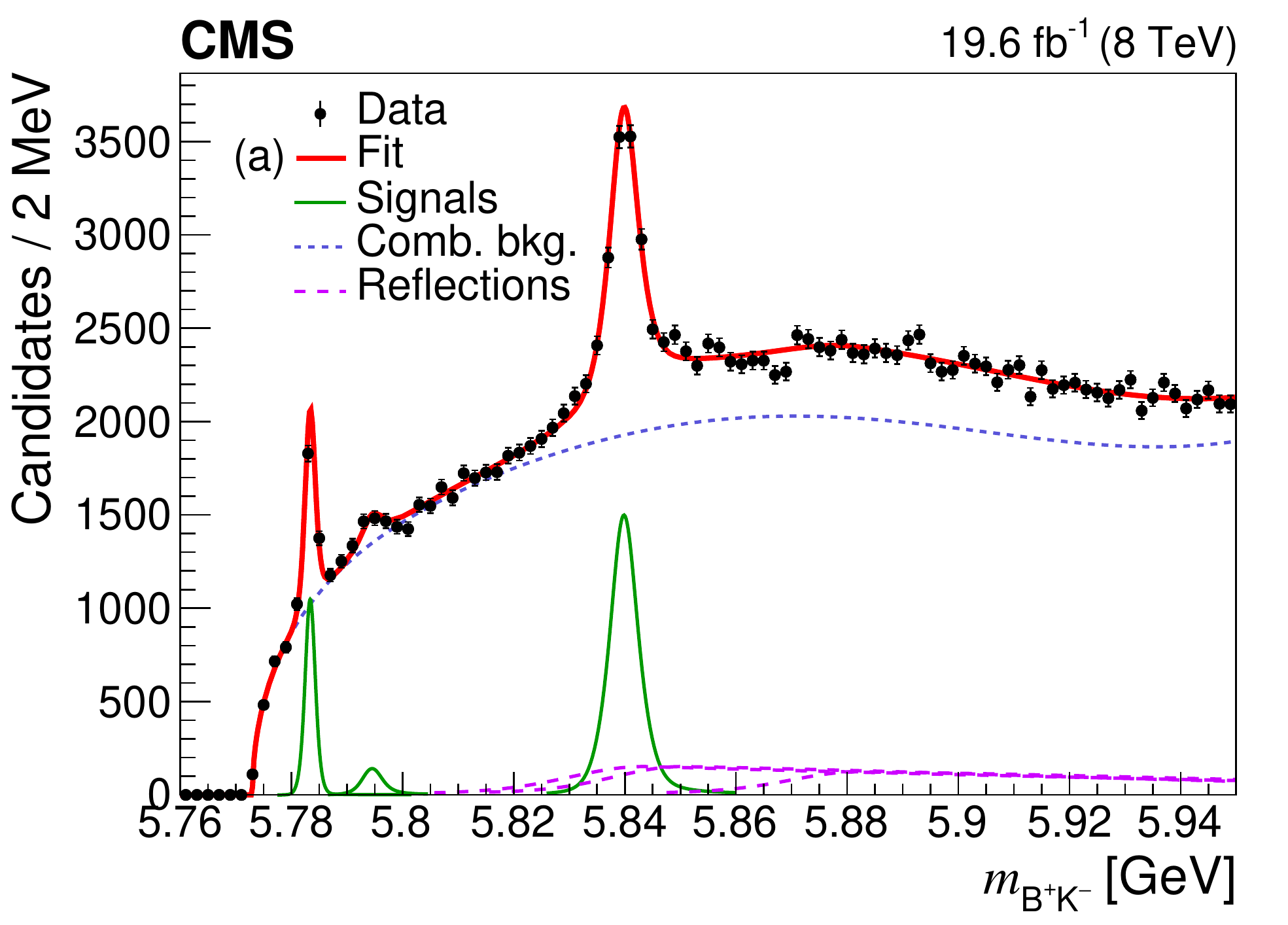}
    \includegraphics[width=0.48\textwidth]{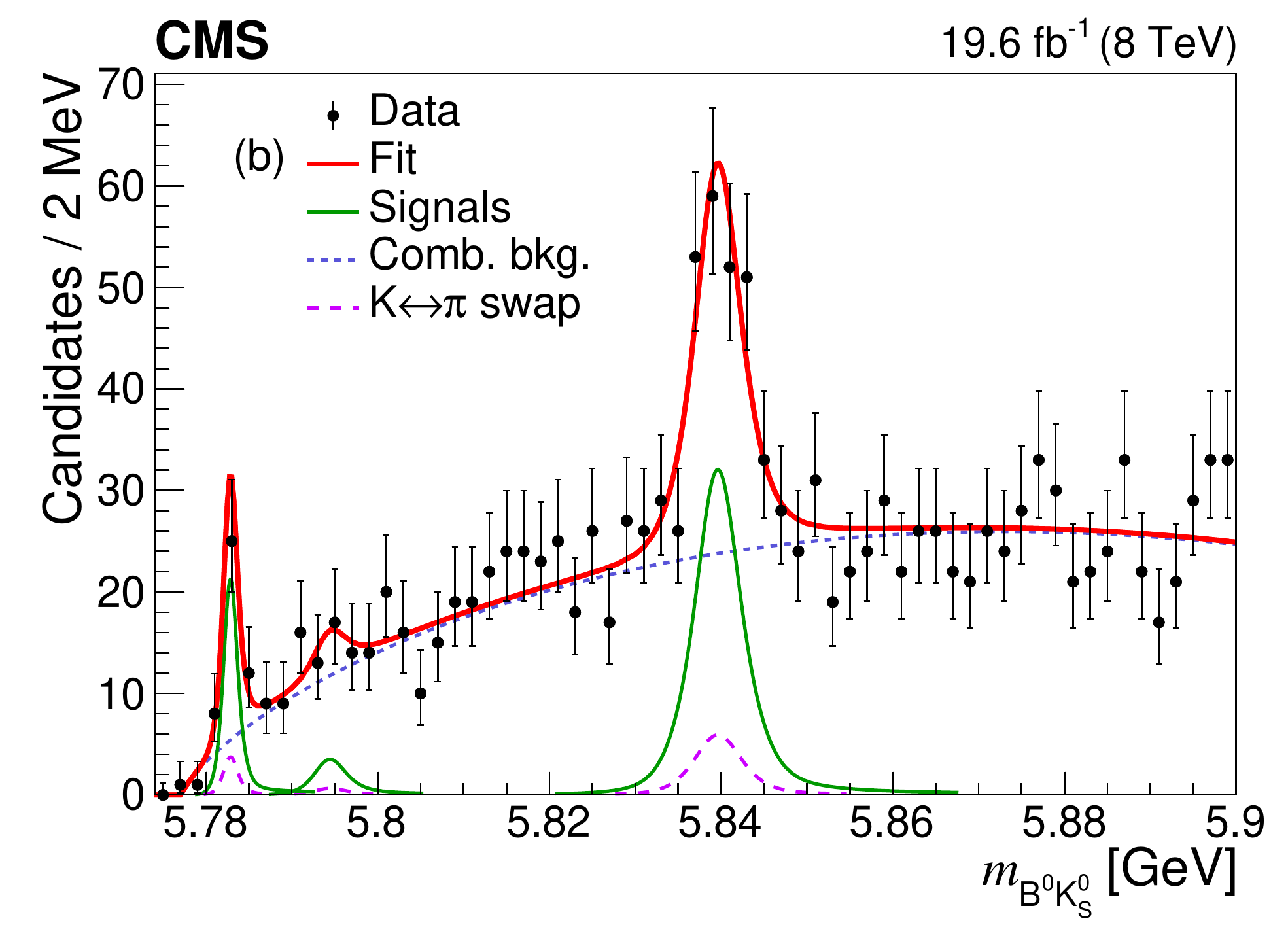}
  \caption{
Invariant mass distributions of (a) $\BuKm$ and (b) $\BdKS$ candidates
with the results of the fit overlaid.
The points represent the data, the thick solid curves are the results of
the overall fits, and the thin solid lines display the signal contributions.
The short-dashed lines show the combinatorial background contributions.
The long-dashed lines show: in (a) the contributions from excited $\Bd$ meson decays,
and in (b) the contributions from swapping $\Kpm\to\pipm$ in
the reconstruction of the $\Bd$ mesons.
\label{fig:mBuKmFits}}
\end{figure}
An unbinned extended maximum-likelihood fit is performed to this
distribution using the sum of three signal functions,
a background function, and the three reflections from the excited $\Bd$ decays.
The signals are described with $D$-wave RBW functions convolved with
double-Gaussian resolution functions obtained from the simulation (the effective
resolutions are about 1--2\MeV).
The natural widths of the $\Bsonetwost$ states and their masses
are free parameters in the fit.
The nonresonant background is modelled by
$(x-x_0)^{\alpha}\, P_n(x)$, where $x\equiv\mBuKm$,
$x_0$ is the threshold value, and the nominal fit uses $n=6$.
The reflections correspond to the contributions of excited $\Bd$ meson
decays into a $\Buorst$ meson and a charged pion, as described above.
The shapes of these contributions are obtained from the simulation
and are fixed in the fit to the data.
The yields of these reflections are corrected by the efficiency of using
the restricted fit region $x_0<\mBuKm<5.95\GeV$.
The results of the fit are presented in the second column of
Table~\ref{tab:mBuKmFfits}, where the measured masses of the $\Bstwost$
and $\Bsone$ mesons are given with respect to the corresponding world-average
$\Bu$ or $\Bust$, and $\Km$ masses~\cite{PDG}.
\par}

\begin{table}[hbt]
  \renewcommand*{\arraystretch}{1.1}
  \centering
  \topcaption{The observed signal yields ($N$), natural widths ($\Gamma$), and mass
  differences from the fits to the $\mBK$ distributions in data.
  The uncertainties are statistical only. \label{tab:mBuKmFfits}}
\cmsTable{
\begin{tabular}{l|cc}
                            & $\BuKm$                   & $\BdKS$                   \\
\hline
$N(\Bstwost\to\B\kaon)$     & $5424\pm269$              & $128\pm22$                \\
$N(\Bstwost\to\Bst\kaon)$   & \phantom{0}$455\pm119$    & \phantom{0}$12\pm11$      \\
$N(\Bsone\to\Bst\kaon)$     & $1329\pm83$\phantom{0}    & $34.5\pm8.3$              \\
[\cmsTabSkip]
$\Gamma(\Bstwost)\,[\MeV]$  & \phantom{0}$1.52\pm0.34$  & \phantom{0}$2.1\pm1.3$    \\
$\Gamma(\Bsone)\,[\MeV]$    & \phantom{0}$0.10\pm0.15$  & \phantom{0}$0.4\pm0.4$    \\
[\cmsTabSkip]
$\dmtwopmzeeq\,[\MeV]$      & $66.93\pm0.09$            & $62.42\pm0.48$            \\
$\dmonepmzeeq\,[\MeV]$      & $10.50\pm0.09$            & \phantom{0}$5.65\pm0.23$  \\
\end{tabular}
}
\end{table}

\subsection{\texorpdfstring{$\BdKS$}{B0 K0S} invariant mass \label{sec:mBdKSfitsdata}}

Similarly to the $\BuKm$ channel, the variable
$\mBdKS = M(\Bd\KS) - M(\Bd) + M_{\Bd}^{\mathrm{PDG}}$
is used for the \BdKS invariant mass. The $\mBdKS$ distribution of the selected $\BdKS$ candidates
is shown in Fig.~\ref{fig:mBuKmFits}(b).
There is a significant peak at about 5840\MeV and a smaller one at 5781\MeV,
corresponding to the decays $\BssBdKS$ and $\BsoBdstKS$, respectively.
The contribution from the $\BssBdstKS$ decay, also shown in
Fig.~\ref{fig:mBuKmFits}(b) at 5795\MeV, is not statistically significant.
However, it is still included in the fit model described below.

The decays $\BssBdKS$, $\BssBdstKS$, and $\BsoBdstKS$ are modelled using
three $D$-wave RBW functions convolved with double-Gaussian resolution functions whose
parameters are fixed according to the simulation.
The masses and natural widths are free parameters in the fit.
Similarly to the $\BuKm$ final state, if the photon from $\Bdst$ decay is lost
and only the $\BdKS$ mass is reconstructed, the peak position is simply shifted
by the mass difference $\mpBdst-\mpBd=45.18\pm0.23\MeV$~\cite{PDG}.
Studies on simulated events show that when the kaon and the pion from
the $\BdJpsiKPi$ decay are exchanged, the three decays
mentioned above produce narrow peaks at the same mass values as the signal peaks.
In order to account for these KPS contributions,
three additional RBW functions, convolved with double-Gaussian shapes, are added,
where the parameters of these Gaussians are fixed to the values obtained in
the simulation and the yields are fixed relative to the signal yields using the
mistagging probability found in the fit to the $\Bd$ invariant mass distribution.
A function of the form $(x-x_0)^{\alpha}\, P_n(x)$ is used to describe the
combinatorial background, where $x\equiv\mBdKS$, $x_0$ is the threshold value,
and $n=1$. The results of the fit are presented in the last column of
Table~\ref{tab:mBuKmFfits}, where the signal yields do not include the KPS component.

The significance of the $\BssBdKS$ decay is estimated
to be 6.3 standard deviations in the baseline fit model using a ratio of the fit
likelihoods with and without the signal component~\cite{Wilks:1938dza}.
Systematic uncertainties, discussed in the next section, are taken into account
using nuisance parameters for the mass resolution, the KPS fraction, and the
$\Bstwost$ mass and natural width. These parameters are allowed to vary
in the fits but are constrained by Gaussian probability density functions.
In particular for the $\Bstwost$ mass and natural width, the world-average values
and their uncertainties~\cite{PDG} are used. Under variations of the fit range
and background model, the significance varies from 6.3 to 7.0 standard deviations.
Similarly, the statistical significance of the $\BsoBdstKS$ signal peak is
3.9 standard deviations, where the systematic uncertainties due to the mass
resolution and KPS fraction are taken into account, as well as the
uncertainties in the $\Bsone$ mass and natural width.
The significance varies from 3.6 to 3.9 standard deviations under
variations of the fit region and the background model.

\section{Efficiencies and systematic uncertainties}

The efficiency for each decay channel is calculated using simulated signal samples.
It is defined as the number of reconstructed signal events from the
simulation divided by the number of generated events. The efficiency includes
the detector acceptance, trigger, and candidate reconstruction efficiencies.
Only the ratios of such efficiencies for different decay modes are needed
in formulae~\eqref{eqON}--\eqref{eqSI}, which reduces the systematic uncertainties
in those ratios. The resulting efficiency ratios used
in the measurements of the ratios of the branching fractions are:
\begin{linenomath}\begin{align*}
\ifthenelse{\boolean{cms@external}}
{ 
\frac{\epsilon(\BssBuKm)}{\epsilon(\BssBdKS)}      &= 15.77\pm0.18,\\
\frac{\epsilon(\BsoBustKm)}{\epsilon(\BsoBdstKS)}  &= 16.33\pm0.20,\\
\frac{\epsilon(\BssBuKm)}{\epsilon(\BssBustKm)}    &= 0.961\pm0.010,\\
\frac{\epsilon(\BssBdKS)}{\epsilon(\BssBdstKS)}    &= 0.970\pm0.012,\\
\frac{\epsilon(\BssBuKm)}{\epsilon(\BsoBustKm)}    &= 0.953\pm0.010,\\
\frac{\epsilon(\BssBdKS)}{\epsilon(\BsoBdstKS)}    &= 0.987\pm0.012,
} 
{ 
\frac{\epsilon(\BssBuKm)}{\epsilon(\BssBdKS)}      &= 15.77\pm0.18,~~~
\frac{\epsilon(\BsoBustKm)}{\epsilon(\BsoBdstKS)}   = 16.33\pm0.20,\\
\frac{\epsilon(\BssBuKm)}{\epsilon(\BssBustKm)}    &= 0.961\pm0.010,~~~
\frac{\epsilon(\BssBdKS)}{\epsilon(\BssBdstKS)}     = 0.970\pm0.012,\\
\frac{\epsilon(\BssBuKm)}{\epsilon(\BsoBustKm)}    &= 0.953\pm0.010,~~~
\frac{\epsilon(\BssBdKS)}{\epsilon(\BsoBdstKS)}     = 0.987\pm0.012,
} 
\end{align*}\end{linenomath}
where the uncertainties are statistical only and related to the
finite size of the simulated samples.

The ratios $\rron$ and $\rrtw$ involve different numbers of
final-state tracks from the decay processes in the numerator and
denominator, and the related signal yields
are extracted from fits to different invariant mass distributions,
unlike the ratios $\rrth$, $\rrfo$, $\rrfi$, and $\rrsi$.
Therefore, the systematic uncertainties are
described separately for the two cases in the next two subsections.

The statistical uncertainties in the efficiency ratios are considered as
sources of systematic uncertainty in the measured branching fraction
ratios. The systematic uncertainties related to muon reconstruction and
identification and trigger efficiencies cancel out in the ratios. Systematic
uncertainties associated with the track reconstruction efficiency are assigned
only in ratios involving final states with a different number of tracks. Validation
studies of the simulated signal samples are performed by comparing distributions
of variables employed in the event selection between simulation and background-subtracted
data, using the channels with the larger yields in data
($\BssBuKm$, $\BsoBustKm$, and $\BssBdKS$).
No significant deviations are found, and no additional systematic uncertainties
in the efficiency ratios are assigned.

\subsection{Systematic uncertainties in the ratios \texorpdfstring{\rron}{R 0 pm 2}
and \texorpdfstring{\rrtw}{R 0 pm 1}\label{systone}}

A systematic uncertainty of $2\times3.9\%=7.8\%$~\cite{Chatrchyan:2014fea} is assigned
to the $\rron$ and $\rrtw$ ratios due to the uncertainty in the track
reconstruction efficiency, since the neutral decay channel has two additional
charged particles in the final state in comparison to the charged decay channel.

{\tolerance=1400
To evaluate the systematic uncertainties related to the choice
of the invariant mass fit model, several alternative functions are tested.
The systematic uncertainty in each signal yield is calculated as
the highest deviation of the observed signal yield from the baseline
fit result. Changes in each fit involve variations in the polynomial degree $n$
in the background model and the fit range; for the fit to the $\mBuPi$ distribution
the variations also include letting the signal masses and natural widths float.
The uncertainties related to fits to the $\BuPi$, $\BuKm$, and $\BdKS$
invariant mass distributions are treated separately and include:
\begin{itemize}
\item A systematic uncertainty related to the fit to $\BuPi$ invariant mass of
2.5\% for $N(\Bstwost\to\Bu\Km)$ and 2.0\% for $N(\Bsone\to\Bust\Km)$,
\item A systematic uncertainty related to the fit to $\BuKm$ invariant mass of
2.4\% for $N(\Bstwost\to\Bu\Km)$ and 4.6\% for $N(\Bsone\to\Bust\Km)$,
\item A systematic uncertainty related to the fit to $\BdKS$ invariant mass of
14\% for $N(\BssBdKS)$ and 8.1\% for $N(\BsoBdstKS)$.
\end{itemize}
\par}

The uncertainty from the invariant mass resolution is estimated by comparing
the $\BuJpsiK$ decays in data and simulation, yielding a difference of at most 2.6\%.
To account for this, the signal fits to the $\mBuKm$ and $\mBdKS$ distributions
in data are repeated with the resolutions decreased and increased by 3\%.
The largest deviations from the baseline in the measured ratios are: 0.7\%
for $N(\BssBdKS)/N(\BssBuKm)$ and 2.2\% for $N(\BsoBdstKS)/N(\BsoBustKm)$.
These values are used as systematic uncertainties in the ratios $\rron$ and $\rrtw$.

The fraction of the KPS component in the $\BdKS$ signals
is obtained from the fit to the $\Bd$ invariant mass distribution in the data.
The systematic uncertainty in this fraction is evaluated by varying the
$\Bd$ signal mass resolution by $\pm3\%$.
The resulting variations of the KPS fraction are at most 3\%.
The other variations in the fit to the $\jpsi\Kstarz$ invariant
mass distribution result in negligible changes in the KPS fraction.
The corresponding systematic uncertainty is 2.6\% in both $\rron$ and $\rrtw$.
As expected, the changes of the
other ratios ($\rrfo$, $\rrsi$) under these variations are negligible.

Formulae~\eqref{eqON} and~\eqref{eqTW} assume the decay $\BdJpsiKPi$
proceeds only through the $\Kstarz$ resonance.
The systematic uncertainty related to this assumption is estimated
by fitting the $\Kp\pim$ invariant mass distribution obtained
from the candidate $\Bd$ data events using the background-subtraction
technique ${}_\mathrm{s}$Plot~\cite{Pivk:2004ty}.
This gives an estimate of 5\% for the nonresonant $\Kp\pim$ fraction
in the total number of signal events, which is included as a
systematic uncertainty in the ratios $\rron$ and $\rrtw$.

All these systematic uncertainties are summarized in
Table~\ref{tab:SYSTone}, along with the total systematic uncertainty,
calculated as the sum in quadrature of the different sources.

\subsection{Systematic uncertainties in the ratios \texorpdfstring{\rrth}{R pm 2*},
\texorpdfstring{\rrfo}{R 0 2*}, \texorpdfstring{\rrfi}{R pm sigma},
and \texorpdfstring{\rrsi}{R 0 sigma}}

\begin{table}[b!]
\renewcommand*{\arraystretch}{1.05}
\centering
  \topcaption{Relative systematic uncertainties in percent in the ratios $\rron$ and $\rrtw$.
  \label{tab:SYSTone}}
  \begin{tabular}{l|cc}
  Source                            & ~~$\rron$~~   &~~$\rrtw$~~\\
    \hline
  Track reconstruction efficiency   & 7.8           & 7.8       \\
  $\mBuPi$ distribution model       & 2.5           & 2.0       \\
  $\mBuKm$ distribution model       & 2.4           & 4.6       \\
  $\mBdKS$ distribution model       & 14            & 8.1       \\
  Mass resolution                   & 0.7           & 2.2       \\
  Fraction of KPS                   & 2.6           & 2.6       \\
  Non-$\Kstarz$ contribution        & 5.0           & 5.0       \\
  Finite size of simulated samples  & 1.2           & 1.2       \\[\cmsTabSkip]
  {Total}                    & {18}   & {14}\\
  \end{tabular}
\renewcommand*{\arraystretch}{1.0}
\end{table}

No systematic uncertainty related to the track reconstruction efficiency is
assigned to the ratios considered in this subsection, since they involve
final states in the numerator and denominator with equal numbers of charged particles.

{\tolerance=1800
In order to evaluate the systematic uncertainties related to the choice of the
invariant mass fit model, several alternative functions are tested, as in the
previous subsection. The systematic uncertainty in each ratio is calculated
as the largest deviation of the corresponding ratio of signal yields obtained
using alternative fit models with respect to the baseline fit model.
The uncertainties related to the fits to $\BuPi$, $\BuKm$, and $\BdKS$
invariant mass distributions are treated separately and include:
\begin{itemize}
\item A systematic uncertainty related to the fit to $\BuPi$ invariant mass of
2.9\% for $N(\BssBustKm)/N(\BssBuKm)$ and
2.7\% for $N(\BsoBustKm)/N(\BssBuKm)$,
\item A systematic uncertainty related to the fit to $\BuKm$ invariant mass of
17\% for $N(\BssBustKm)/N(\BssBuKm)$ and
7.1\% for $N(\BsoBustKm)/N(\BssBuKm)$,
\item A systematic uncertainty related to the fit to $\BdKS$ invariant mass of
13\% for $N(\BssBdstKS)/N(\BssBdKS)$ and
24\% for the ratio $N(\BsoBdstKS)/N(\BssBdKS)$.
\end{itemize}
\par}

The systematic uncertainty in the ratios $\rrth$, $\rrfo$, $\rrfi$, and $\rrsi$,
related to the knowledge of the invariant mass resolution is estimated as in
the previous subsection, and is found to be in the range 1.2--3.0\%.

{\tolerance=800
The systematic uncertainty associated with the uncertainty in the mass differences
$\mpBust-\mpBu$ and $\mpBdst-\mpBd$ must be taken into account, since these
values are fixed in the fits. The baseline fits are repeated with
each mass difference fixed to its nominal value plus and minus its uncertainty,
and the largest deviations from the baseline of the obtained ratios of signal
yields are taken as systematic uncertainties:
7.7\% for $N(\BssBustKm)/N(\BssBuKm)$ and
4.8\% for $N(\BssBdstKS)/N(\BssBdKS)$.
The changes in other ratios under variations of $\mpBust-\mpBu$ and $\mpBdst-\mpBd$
are negligible.
\par}

The systematic uncertainties due to non-$\Kstarz$ contributions cancel out in
the ratios $\rrfo$ and $\rrsi$.

Table~\ref{tab:SYSTtwo} lists those systematic uncertainties, together with the
total ones, calculated by summing the different contributions in quadrature.
\begin{table*}[h!]
\renewcommand*{\arraystretch}{1.05}
\centering
  \topcaption{Relative systematic uncertainties in percent in the ratios
  $\rrth$, $\rrfo$, $\rrfi$, and $\rrsi$. \label{tab:SYSTtwo}}
  \begin{tabular}{l|cccc}
  Source                            & ~~$\rrth$~~   &~~$\rrfo$~~    &~~$\rrfi$~~    &~~$\rrsi$~~\\
    \hline
  $\mBuPi$ distribution model       & 2.9           & \NA           & 2.7           & \NA       \\
  $\mBuKm$ distribution model       & 17            & \NA           & 7.1           & \NA       \\
  $\mBdKS$ distribution model       & \NA           & 13            & \NA           & 24        \\
  Mass resolution                   & 1.2           & 3.0           & 1.5           & 1.1       \\
  Uncertainties in $\mpBst-\mpB$    & 7.7           & 4.8           & \NA           & \NA       \\
  Finite size of simulated samples  & 1.1           & 1.3           & 1.1           & 1.3       \\[\cmsTabSkip]
  {Total}                    & {19}   & {15}   & {7.8}   & {24}
  \end{tabular}
\renewcommand*{\arraystretch}{1.0}
\end{table*}

\subsection{Systematic uncertainties in the mass differences and
natural widths\label{sec:SystMG}}

The fits to the $\BK$ invariant mass distributions are also used
to measure the mass differences
\begin{linenomath}\begin{equation*}
\ifthenelse{\boolean{cms@external}}
{ 
\begin{aligned}
\dmtwopm &= \dmtwopmeq, \\
\dmonepm &= \dmonepmeq, \\
\dmtwoze &= \dmtwozeeq, \\
\dmoneze &= \dmonezeeq.
\end{aligned}
} 
{ 
\begin{aligned}
\dmtwopm &= \dmtwopmeq, & \dmonepm &= \dmonepmeq,\\
\dmtwoze &= \dmtwozeeq, & \dmoneze &= \dmonezeeq.
\end{aligned}
} 
\end{equation*}\end{linenomath}
Using these values, the mass differences
\begin{linenomath}
\ifthenelse{\boolean{cms@external}}
{ 
\begin{align*}
\mBd-\mBu &= \dmtwopm-\dmtwoze+M_{\Km}^{\mathrm{PDG}}-M_{\KS}^{\mathrm{PDG}}\\
\intertext{and}
\mBdst-\mBust &= \dmonepm-\dmoneze+M_{\Km}^{\mathrm{PDG}}-M_{\KS}^{\mathrm{PDG}}
\end{align*}
} 
{ 
\begin{align*}
\mBd-\mBu &= \dmtwopm-\dmtwoze+M_{\Km}^{\mathrm{PDG}}-M_{\KS}^{\mathrm{PDG}}\\
\intertext{and}
\mBdst-\mBust &= \dmonepm-\dmoneze+M_{\Km}^{\mathrm{PDG}}-M_{\KS}^{\mathrm{PDG}}
\end{align*}
} 
\end{linenomath}
can be determined.

The natural width of the $\Bstwost$ state is measured only in the $\BuKm$ channel
due to the limited number of events in the $\BdKS$ channel. Systematic uncertainties in
these measurements are discussed in this subsection.

The uncertainty related to the choice of the fit model is estimated by
testing alternative fit models, as in Section~\ref{systone}.
The largest deviation from the mass difference obtained from each baseline
fit value is taken as the systematic uncertainty in the respective mass
difference or natural width.
The uncertainties related to the fits to the $\BuPi$, $\BuKm$, and $\BdKS$
invariant mass distributions are treated separately.

The systematic uncertainty associated with the knowledge of the mass difference
$\mpBust-\mpBu$ (or $\mpBdst-\mpBd$) is taken into account as well: the baseline fits
are repeated with the mass difference $\mpBst-\mpB$ fixed to its nominal value plus or
minus its uncertainty. The largest deviation from the baseline of the obtained
mass differences and natural width
is taken as the corresponding systematic uncertainty.

Studies of simulated events show that the mass differences measured in the
reconstructed invariant mass distributions are slightly shifted with respect
to the mass differences used in the generation of simulated events. Therefore,
the measured mass differences are corrected by the observed shifts (which
are up to 0.056\MeV), and each shift is conservatively
treated as a systematic uncertainty in the respective mass-difference measurement.

In order to estimate the systematic uncertainties due to possible
misalignment of the detector~\cite{Chatrchyan:2014wfa}, eighteen different
simulated samples with various distorted geometries are
produced and analyzed for each of the four decay channels.
From these measurements the largest deviation of the
estimation of the invariant mass or its resolution from the
perfectly aligned case is accepted as an estimate of the
systematic uncertainty from a possible detector misalignment.
The magnitudes of distortions are large enough to be detected and corrected
by the standard alignment procedures~\cite{Chatrchyan:2014wfa}.
The shifts in the measured mass differences
observed in these simulations are up to 0.038\MeV. The
systematic uncertainty in the invariant mass resolution of the $\BssBuKm$ signal
is found to be 0.042\MeV, and the corresponding uncertainty in $\Gammabss$ is
obtained by repeating the baseline fit with the resolution increased or
decreased by this value.
The largest deviation in the measured natural width with respect to the baseline value
is used as a systematic uncertainty.

The systematic uncertainties related to the invariant mass resolution are estimated
in the same way as in the previous subsections and are found to be up to 0.007\MeV
for the mass differences and 0.2\MeV for the natural width. This source of
uncertainty is conservatively considered to be uncorrelated with the systematic
uncertainty related to a possible detector misalignment.

These systematic uncertainties are summarized in
Table~\ref{tab:SYSTthree}, together with the total systematic uncertainties,
calculated by summing in quadrature the different contributions.
It was checked that the mass of the \Bu meson, measured in the
$\BuJpsiK$ decay, is consistent with the world-average value, after taking
into account the systematic uncertainties related to the shift from
the reconstruction and possible detector misalignment.

\begin{table*}[h!]
\renewcommand*{\arraystretch}{1.05}
\centering
  \topcaption{Systematic uncertainties (in \MeV) in the measured
  mass differences and natural width. The $\Bstwost$ width is
  measured only in the $\BuKm$ channel. \label{tab:SYSTthree}}
\tableCMS{
  \begin{tabular}{l|ccccccc}
\rule{0pt}{13pt} {Source}                   &~$\dmtwopm$    &~$\dmonepm$    &~$\dmtwoze$    &~$\dmoneze$    &$\mBd-\mBu$    &$\mBdst-\mBust$&~$\Gammabss$\\[4pt]
    \hline
  $\mBuPi$ distribution model               & 0.024         & 0.008         & \NA           & \NA           &  0.024        &   0.008       & 0.11      \\
  $\mBuKm$ distribution model               & 0.011         & 0.043         & \NA           & \NA           &  0.011        &   0.043       & 0.11      \\
  $\mBdKS$ distribution model               & \NA           & \NA           & 0.039         & 0.038         &  0.039        &   0.038       & \NA       \\
  Uncertainties in $\mpBst-\mpB$            & 0.012         & 0.003         & 0.003         & 0.0001        &  0.012        &   0.003       & 0.03      \\
  Shift from reconstruction                 & 0.056         & 0.044         & 0.050         & 0.042         &  0.075        &   0.061       & \NA       \\
  Detector misalignment                     & 0.036         & 0.005         & 0.031         & 0.006         &  0.038        &   0.008       & 0.15      \\
  Mass resolution                           & 0.007         & 0.005         & 0.005         & 0.005         &  0.009        &   0.007       & 0.20      \\[\cmsTabSkip]
  {Total}                            &{0.073} &{0.063}  &{0.071}&{0.057} &{0.098} &{0.085} &{0.30}
  \end{tabular}
}
\renewcommand*{\arraystretch}{1.0}
\end{table*}

\section{Results}
The decay $\BssBdKS$ is observed for the
first time with a corresponding statistical significance of 6.3 standard deviations.
The first evidence (3.9 standard deviations) for the decay $\BsoBdstKS$ is found.
In the measurements presented below of the relative branching fractions,
cross sections multiplied by branching fractions, masses, mass differences,
and natural width, the first uncertainty is statistical, the second is systematic,
and if there is a third, it is related to the uncertainties in the
world-average values of the branching fractions, masses, and mass differences~\cite{PDG}.

{\tolerance=800
Formulae~\eqref{eqON}--\eqref{eqFO} are used with the branching
fractions~\cite{PDG}
$\mathcal{B}(\BuJpsiK)=(1.026\pm0.031)\,10^{-3},$
$\mathcal{B}(\BdJpsiKst)=(1.28\pm0.05)\,10^{-3},$
$\mathcal{B}(\KstarKPi)=(0.99754\pm0.00021),$ and
$\mathcal{B}(\KSPiPi)=(0.6920\pm0.0005)$
to determine the following ratios of branching fractions:
\begin{linenomath}
\ifthenelse{\boolean{cms@external}}
{ 
\begin{align*}
\rron &= \frac{\mathcal{B}(\BssBdKS)}{\mathcal{B}(\BssBuKm)}     = 0.432\pm0.077\pm0.075\pm0.021, \\
\rrtw &= \frac{\mathcal{B}(\BsoBdstKS)}{\mathcal{B}(\BsoBustKm)} = 0.49\pm0.12\pm0.07\pm0.02, \\
\rrth &= \frac{\mathcal{B}(\BssBustKm)}{\mathcal{B}(\BssBuKm)}   = 0.081\pm0.021\pm0.015, \\
\rrfo &= \frac{\mathcal{B}(\BssBdstKS)}{\mathcal{B}(\BssBdKS)}   = 0.093\pm0.086\pm0.014.
\end{align*}
} 
{ 
\begin{alignat*}{2}
\rron &= \frac{\mathcal{B}(\BssBdKS)}{\mathcal{B}(\BssBuKm)}     &&= 0.432\pm0.077\pm0.075\pm0.021, \\
\rrtw &= \frac{\mathcal{B}(\BsoBdstKS)}{\mathcal{B}(\BsoBustKm)} &&= 0.49\pm0.12\pm0.07\pm0.02, \\
\rrth &= \frac{\mathcal{B}(\BssBustKm)}{\mathcal{B}(\BssBuKm)}   &&= 0.081\pm0.021\pm0.015, \\
\rrfo &= \frac{\mathcal{B}(\BssBdstKS)}{\mathcal{B}(\BssBdKS)}   &&= 0.093\pm0.086\pm0.014.
\end{alignat*}
} 
\end{linenomath}
The ratio $\rron$ is in good agreement with the theoretical predictions
of about 0.43~\cite{Wang:2012pf, Lu:2016bbk}, while the ratio $\rrtw$
is $2.5$ standard deviations away from the theoretical prediction of
0.23~\cite{Wang:2012pf}, which, however, has no uncertainty estimate.
The third ratio is in agreement with the measurements of LHCb~\cite{Aaij:2012uva}
and CDF~\cite{Aaltonen:2013atp}: $0.093\pm0.013\pm0.012$ and
$0.10\pm0.03\pm0.02$, respectively. It is also consistent with
the theoretical predictions~\cite{Zhong:2008kd, Colangelo:2012xi, Wang:2012pf, Lu:2016bbk}.
The fourth ratio is a new result.
\par}

In addition, using Eqs.~\eqref{eqFI}--\eqref{eqSI}, the ratios of
production cross sections times branching fractions are measured:
\begin{linenomath}\begin{equation*}
\ifthenelse{\boolean{cms@external}}
{ 
\begin{aligned}
\rrfi &= \frac{\sigma(\pp\to\Bsone \mathrm{X})\,\mathcal{B}(\BsoBustKm)}{\sigma(\pp\to\Bstwost \mathrm{X})\,\mathcal{B}(\BssBuKm)} \\ &=
0.233\pm0.019\pm0.018, \\
\rrsi &= \frac{\sigma(\pp\to\Bsone \mathrm{X})\,\mathcal{B}(\BsoBdstKS)}{\sigma(\pp\to\Bstwost \mathrm{X})\,\mathcal{B}(\BssBdKS)} \\ &=
0.266\pm0.079\pm0.063.
\end{aligned}
} 
{ 
\begin{aligned}
\rrfi = \frac{\sigma(\pp\to\Bsone \mathrm{X})\,\mathcal{B}(\BsoBustKm)}{\sigma(\pp\to\Bstwost \mathrm{X})\,\mathcal{B}(\BssBuKm)} &=
0.233\pm0.019\pm0.018, \\
\rrsi = \frac{\sigma(\pp\to\Bsone \mathrm{X})\,\mathcal{B}(\BsoBdstKS)}{\sigma(\pp\to\Bstwost \mathrm{X})\,\mathcal{B}(\BssBdKS)} &=
0.266\pm0.079\pm0.063.
\end{aligned}
} 
\end{equation*}\end{linenomath}
The value of $\rrfi$ was previously determined by LHCb to be
$0.232\pm0.014\pm0.013$~\cite{Aaij:2012uva} at $\sqrt{s}=7\TeV$ and
in a different pseudorapidity region, consistent with the result
presented here.

The following mass differences are obtained:
\begin{linenomath}\begin{equation*}
\ifthenelse{\boolean{cms@external}}
{ 
\begin{aligned}
\dmtwopm &= \dmtwopmeq \\ &= 66.87\pm0.09\pm0.07\MeV, \\
\dmtwoze &= \dmtwozeeq \\ &= 62.37\pm0.48\pm0.07\MeV, \\
\dmonepm &= \dmonepmeq \\ &= 10.45\pm0.09\pm0.06\MeV, \\
\dmoneze &= \dmonezeeq \\ &=  5.61\pm0.23\pm0.06\MeV.
\end{aligned}
} 
{ 
\begin{aligned}
\dmtwopm = \dmtwopmeq &= 66.87\pm0.09\pm0.07\MeV, \\
\dmtwoze = \dmtwozeeq &= 62.37\pm0.48\pm0.07\MeV, \\
\dmonepm = \dmonepmeq &= 10.45\pm0.09\pm0.06\MeV, \\
\dmoneze = \dmonezeeq &=  5.61\pm0.23\pm0.06\MeV.
\end{aligned}
} 
\end{equation*}\end{linenomath}
The first two mass differences are in good agreement
with LHCb~\cite{Aaij:2012uva} and CDF~\cite{Aaltonen:2013atp} results
(see Table~\ref{tab:PreviousResults}).
Using these two measurements, the world-average masses of the $\Bu$ and $\Km$ mesons,
and the mass difference $\mpBust-\mpBu$, the $\Bsonetwost$ masses are determined:
\begin{linenomath}\begin{equation*}\begin{aligned}
M(\Bstwost) &= 5839.86\pm0.09\pm0.07\pm0.15\MeV,\\
M(\Bsone)   &= 5828.78\pm0.09\pm0.06\pm0.28\MeV.\\
\end{aligned}\end{equation*}\end{linenomath}
The measured masses in the $\BdKS$ channel are consistent with
our results using the $\BuKm$ channel but have significantly larger uncertainties.

{\tolerance=800
Using the mass-difference measurements above, the mass differences between the neutral
and charged $\B$ and $\Bst$ mesons are found to be:
\begin{linenomath}\begin{equation*}\begin{aligned}
\mBd-\mBu       &= 0.57\pm0.49\pm0.10\pm0.02\MeV, \\
\mBdst-\mBust   &= 0.91\pm0.24\pm0.09\pm0.02\MeV.
\end{aligned}\end{equation*}\end{linenomath}
The first mass difference result is consistent with the significantly more
precise world-average value of $0.31\pm0.06\MeV$~\cite{PDG}.
There are no previous measurements of $\mBdst-\mBust$, and this paper
presents a new method to measure both of these mass differences.
\par}

Lastly, the natural width of the $\Bstwost$ meson is determined to be
\begin{linenomath}\begin{equation*}
\Gammabss= 1.52\pm0.34\pm0.30\MeV,
\end{equation*}\end{linenomath}
consistent with the results of LHCb~\cite{Aaij:2012uva}
and CDF~\cite{Aaltonen:2013atp} (see Table~\ref{tab:PreviousResults}).

\section{Summary}
{\tolerance=800
The $P$-wave $\Bs$ meson states are studied using a data
sample corresponding to an integrated luminosity of
$19.6\fbinv$ of proton-proton collisions collected by the CMS experiment
at $\sqrt{s}= 8\TeV$ in 2012. Observation and evidence are reported
for the decays $\BssmBdKS$ and $\BsomBdstKS$, respectively.
Four ratios of branching fractions and two ratios of production cross
sections multiplied by the branching fractions of the $P$-wave $\Bs$ mesons
into a $\PB$ meson and kaon are measured. In addition, the differences between the
$\Bsonetwost$ mass and the sum of the $\PB$ meson and kaon mass are determined,
as well as the $\Bstwostm$ natural width.
Finally, using a new approach, the mass differences $\mBd-\mBu$ and
$\mBdst-\mBust$ are measured, where the latter is determined for the first time.
\par}

\begin{acknowledgments} 
\hyphenation{Bundes-ministerium Forschungs-gemeinschaft Forschungs-zentren Rachada-pisek} We congratulate our colleagues in the CERN accelerator departments for the excellent performance of the LHC and thank the technical and administrative staffs at CERN and at other CMS institutes for their contributions to the success of the CMS effort. In addition, we gratefully acknowledge the computing centres and personnel of the Worldwide LHC Computing Grid for delivering so effectively the computing infrastructure essential to our analyses. Finally, we acknowledge the enduring support for the construction and operation of the LHC and the CMS detector provided by the following funding agencies: the Austrian Federal Ministry of Education, Science and Research and the Austrian Science Fund; the Belgian Fonds de la Recherche Scientifique, and Fonds voor Wetenschappelijk Onderzoek; the Brazilian Funding Agencies (CNPq, CAPES, FAPERJ, FAPERGS, and FAPESP); the Bulgarian Ministry of Education and Science; CERN; the Chinese Academy of Sciences, Ministry of Science and Technology, and National Natural Science Foundation of China; the Colombian Funding Agency (COLCIENCIAS); the Croatian Ministry of Science, Education and Sport, and the Croatian Science Foundation; the Research Promotion Foundation, Cyprus; the Secretariat for Higher Education, Science, Technology and Innovation, Ecuador; the Ministry of Education and Research, Estonian Research Council via IUT23-4 and IUT23-6 and European Regional Development Fund, Estonia; the Academy of Finland, Finnish Ministry of Education and Culture, and Helsinki Institute of Physics; the Institut National de Physique Nucl\'eaire et de Physique des Particules~/~CNRS, and Commissariat \`a l'\'Energie Atomique et aux \'Energies Alternatives~/~CEA, France; the Bundesministerium f\"ur Bildung und Forschung, Deutsche Forschungsgemeinschaft, and Helmholtz-Gemeinschaft Deutscher Forschungszentren, Germany; the General Secretariat for Research and Technology, Greece; the National Research, Development and Innovation Fund, Hungary; the Department of Atomic Energy and the Department of Science and Technology, India; the Institute for Studies in Theoretical Physics and Mathematics, Iran; the Science Foundation, Ireland; the Istituto Nazionale di Fisica Nucleare, Italy; the Ministry of Science, ICT and Future Planning, and National Research Foundation (NRF), Republic of Korea; the Ministry of Education and Science of the Republic of Latvia; the Lithuanian Academy of Sciences; the Ministry of Education, and University of Malaya (Malaysia); the Ministry of Science of Montenegro; the Mexican Funding Agencies (BUAP, CINVESTAV, CONACYT, LNS, SEP, and UASLP-FAI); the Ministry of Business, Innovation and Employment, New Zealand; the Pakistan Atomic Energy Commission; the Ministry of Science and Higher Education and the National Science Centre, Poland; the Funda\c{c}\~ao para a Ci\^encia e a Tecnologia, Portugal; JINR, Dubna; the Ministry of Education and Science of the Russian Federation, the Federal Agency of Atomic Energy of the Russian Federation, Russian Academy of Sciences, the Russian Foundation for Basic Research, and the National Research Center ``Kurchatov Institute"; the Ministry of Education, Science and Technological Development of Serbia; the Secretar\'{\i}a de Estado de Investigaci\'on, Desarrollo e Innovaci\'on, Programa Consolider-Ingenio 2010, Plan Estatal de Investigaci\'on Cient\'{\i}fica y T\'ecnica y de Innovaci\'on 2013-2016, Plan de Ciencia, Tecnolog\'{i}a e Innovaci\'on 2013-2017 del Principado de Asturias, and Fondo Europeo de Desarrollo Regional, Spain; the Ministry of Science, Technology and Research, Sri Lanka; the Swiss Funding Agencies (ETH Board, ETH Zurich, PSI, SNF, UniZH, Canton Zurich, and SER); the Ministry of Science and Technology, Taipei; the Thailand Center of Excellence in Physics, the Institute for the Promotion of Teaching Science and Technology of Thailand, Special Task Force for Activating Research and the National Science and Technology Development Agency of Thailand; the Scientific and Technical Research Council of Turkey, and Turkish Atomic Energy Authority; the National Academy of Sciences of Ukraine, and State Fund for Fundamental Researches, Ukraine; the Science and Technology Facilities Council, UK; the US Department of Energy, and the US National Science Foundation.

Individuals have received support from the Marie-Curie programme and the European Research Council and Horizon 2020 Grant, contract No. 675440 (European Union); the Leventis Foundation; the A. P. Sloan Foundation; the Alexander von Humboldt Foundation; the Belgian Federal Science Policy Office; the Fonds pour la Formation \`a la Recherche dans l'Industrie et dans l'Agriculture (FRIA-Belgium); the Agentschap voor Innovatie door Wetenschap en Technologie (IWT-Belgium); the F.R.S.-FNRS and FWO (Belgium) under the ``Excellence of Science - EOS" - be.h project n. 30820817; the Ministry of Education, Youth and Sports (MEYS) of the Czech Republic; the Lend\"ulet (``Momentum") Programme and the J\'anos Bolyai Research Scholarship of the Hungarian Academy of Sciences, the New National Excellence Program \'UNKP, the NKFIA research grants 123842, 123959, 124845, 124850 and 125105 (Hungary); the Council of Scientific and Industrial Research, India; the HOMING PLUS programme of the Foundation for Polish Science, cofinanced from European Union, Regional Development Fund, the Mobility Plus programme of the Ministry of Science and Higher Education, the National Science Center (Poland), contracts Harmonia 2014/14/M/ST2/00428, Opus 2014/13/B/ST2/02543, 2014/15/B/ST2/03998, and 2015/19/B/ST2/02861, Sonata-bis 2012/07/E/ST2/01406; the Ministry of Education and Science of the Russian Federation contract No. 14.W03.31.0026; the National Priorities Research Program by Qatar National Research Fund; the Programa de Excelencia Mar\'{i}a de Maeztu, and the Programa Severo Ochoa del Principado de Asturias; the Thalis and Aristeia programmes cofinanced by EU-ESF, and the Greek NSRF; the Rachadapisek Sompot Fund for Postdoctoral Fellowship, Chulalongkorn University, and the Chulalongkorn Academic into Its 2nd Century Project Advancement Project (Thailand); the Welch Foundation, contract C-1845; and the Weston Havens Foundation (USA).
\end{acknowledgments}

\bibliography{auto_generated}

\cleardoublepage \appendix\section{The CMS Collaboration \label{app:collab}}\begin{sloppypar}\hyphenpenalty=5000\widowpenalty=500\clubpenalty=5000\input{BPH-16-003-authorlist.tex}\end{sloppypar}
\end{document}

%% file: BPH-16-003-authorlist.tex
\vskip\cmsinstskip
\textbf{Yerevan Physics Institute, Yerevan, Armenia}\\*[0pt]
A.M.~Sirunyan, A.~Tumasyan
\vskip\cmsinstskip
\textbf{Institut f\"{u}r Hochenergiephysik, Wien, Austria}\\*[0pt]
W.~Adam, F.~Ambrogi, E.~Asilar, T.~Bergauer, J.~Brandstetter, M.~Dragicevic, J.~Er\"{o}, A.~Escalante~Del~Valle, M.~Flechl, R.~Fr\"{u}hwirth\cmsAuthorMark{1}, V.M.~Ghete, J.~Hrubec, M.~Jeitler\cmsAuthorMark{1}, N.~Krammer, I.~Kr\"{a}tschmer, D.~Liko, T.~Madlener, I.~Mikulec, N.~Rad, H.~Rohringer, J.~Schieck\cmsAuthorMark{1}, R.~Sch\"{o}fbeck, M.~Spanring, D.~Spitzbart, A.~Taurok, W.~Waltenberger, J.~Wittmann, C.-E.~Wulz\cmsAuthorMark{1}, M.~Zarucki
% ...
\vskip\cmsinstskip
% ...
1:  Also at Vienna University of Technology, Vienna, Austria\\
% ...
